\documentclass[dvipsnames,twocolumn]{aastex701}

\usepackage[T1]{fontenc}
\usepackage{nicefrac}
\usepackage{savesym}
\savesymbol{tablenum}
\usepackage{siunitx}
\usepackage{amsmath}
\newcommand{\RN}[1]{%
  \textup{\uppercase\expandafter{\romannumeral#1}}%
}
\usepackage{mathtools}
\usepackage{booktabs}
\usepackage{multirow}
\usepackage[normalem]{ulem}
\usepackage{soul}
\usepackage{natbib}
\usepackage{dcolumn}
\usepackage{xcolor}
\usepackage{tikz}
\usetikzlibrary{shapes.geometric, arrows}
\usepackage{cleveref}
\usepackage{CJK}
\restoresymbol{SIX}{tablenum}
\DeclareSIUnit\parsec{pc}
\DeclareSIUnit\years{yr}
\DeclareSIUnit\Msol{M_{\odot}}
\DeclareSIUnit\Lsol{L_{\odot}}
\DeclareSIUnit\erg{erg}
\DeclareSIUnit\Mearth{M_{\earth}}
\DeclareSIUnit\Mjup{M_{\mathrm{Jup}}}
\DeclareSIUnit\Mth{M_{\mathrm{th}}}
\DeclareSIUnit\Lsol{L_{\odot}}
\DeclareSIUnit\AU{au}
\DeclareSIUnit\om{\Omega}
\DeclareSIUnit\orb{T_{\mathrm{orb}}}
\DeclareSIUnit\scaleheight{H}
\definecolor{dodgerblue}{rgb}{0.11764706, 0.56470588, 1.}
\definecolor{seagreen}{rgb}{0.18039216, 0.54509804, 0.34117647}
\definecolor{maroon}{rgb}{0.50196078, 0., 0.}

\shortauthors{Pfeil et al.}

\newcommand{\rev}[1]{{{#1}}}
\newcommand{\revII}[1]{{{#1}}}
\newcommand{\rem}[1]{{{}}}

\newcommand{\Diff}[2]{\frac{\mathrm{d}{#1}}{\mathrm{d}{#2}}}
\newcommand{\diff}[2]{\frac{\partial{#1}}{\partial{#2}}}
\newcommand{\vect}[1]{\ensuremath{\mathbf{#1}}}
\newcommand{\rhog}{\Sigma_{\mathrm{g}}}

\newcommand{\rhodn}{\Sigma_{\mathrm{d},n}}
\newcommand{\rhodsm}{\Sigma_{\mathrm{d},0}}
\newcommand{\rhodlr}{\Sigma_{\mathrm{d},1}}
\newcommand{\vg}{\vect{v}_{\mathrm{g}}}

\newcommand{\vdn}{\vect{v}_{\mathrm{d},n}}
\newcommand{\amax}{a_{\mathrm{max}}}
\newcommand{\amin}{a_{\mathrm{min}}}
\newcommand{\aint}{a_{\mathrm{int}}}

\newcommand{\St}{\mathrm{St}}

\newcommand{\dpy}{{\normalfont\texttt{DustPy}}}
\newcommand{\athena}{{\normalfont\texttt{Athena++}}}
\newcommand{\PLUTO}{{\normalfont\texttt{PLUTO}}}
\newcommand{\LACOMPASS}{{\normalfont\texttt{LA-COMPASS}}}

\newcommand{\tpop}{{\normalfont\texttt{TriPoD}}}

\begin{document}
\begin{CJK*}{UTF8}{gbsn}

\title{Dust fragmentation enhances the leakage of dust through planetary gaps. \\ Implications for Solar System compositional dichotomies and the dust mass in the inner regions of protoplanetary disk.}

\title{\rem{Dust fragmentation enhances the leakage of dust through planetary gaps. \\ Implications for Solar System compositional dichotomies and the dust mass in the inner regions of protoplanetary disk.}

Fragmentation-limited dust filtration in 2D simulations of planet-disk systems with dust coagulation. \\
Parameter study and implications for the inner disk's dust mass budget and composition.}

\correspondingauthor{Thomas Pfeil}
\author[0000-0002-4171-7302]{Thomas Pfeil}
\affiliation{Center for Computational Astrophysics, Flatiron Institute, 162 Fifth Avenue, New York, NY 10010, USA\footnote{The Flatiron Institute is a division of the Simons Foundation.}}
\email{tpfeil@flatironinstitute.org}

\author[0000-0001-5032-1396]{Philip J. Armitage}
\affiliation{Center for Computational Astrophysics, Flatiron Institute, 162 Fifth Avenue, New York, NY 10010, USA\footnote{The Flatiron Institute is a division of the Simons Foundation.}}
\affiliation{Department of Physics and Astronomy, Stony Brook University, Stony Brook, NY 11794, USA}
\email{philip.armitage@stonybrook.edu}

\author[0000-0002-2624-3399]{Yan-Fei Jiang (姜燕飞)}
\affiliation{Center for Computational Astrophysics, Flatiron Institute, 162 Fifth Avenue, New York, NY 10010, USA\footnote{The Flatiron Institute is a division of the Simons Foundation.}}
\email{yjiang@flatironinstitute.org}

\begin{abstract}
\rem{
Super-thermal gas giant planets or their progenitor cores are known to open deep gaps in protoplanetary disks, which stop large, drifting dust particles on their way to the inner disk. The possible separation of the disk into distinct reservoirs and the resulting dust depletion interior to the gap have important implications for planetesimal formation and the chemical and isotopic composition of the inner regions of protoplanetary disks.
Small grains, however, can diffuse through a gap and mostly follow the gas flow. Coagulation and fragmentation determine the available mass of small grains and are thus instrumental for the study of a gap's filtration efficiency. We present two-dimensional multifluid hydrodynamic simulations of planet-disk systems with dust coagulation, evolved over \SI{45000} planetary orbits, to investigate the effects of different planetary masses, dust fragmentation velocities, and viscosities on the inner disk's dust mass budget and composition. We find that filtering can only be efficient for high planetary masses, high fragmentation velocities, and low diffusivities. Clear compositional distinctions between the inner and outer disk cannot be maintained by a \SI{31}{\Mearth} planet if the fragmentation velocity is low, even if $\alpha \lesssim 5\times 10^{-4}$. Significant ``contamination'' of the inner disk by outer disk dust occurs in much less than \SI{2e5}{\years} for Jupiter's core and even for more massive objects. This either places tight constraints on the physical conditions in the Solar nebula or mandates consideration of alternative explanations for the NC-CC dichotomy. Astrophysical constraints on the parameters could discriminate between these possibilities.\newline
}
Super-thermal gas giant planets or their progenitor cores are known to open deep gaps in protoplanetary disks, which stop large, drifting dust particles on their way to the inner disk. The possible separation of the disk into distinct reservoirs and the resulting dust depletion interior to the gap have important implications for planetesimal formation and the chemical and isotopic composition of the inner regions of protoplanetary disks.
Dust fragmentation, however, maintains a reservoir of small grains which can traverse the gap. Dust evolution models are thus instrumental for studies of a gap's filtration efficiency. We present 2D multifluid hydrodynamic simulations of planet-disk systems with dust coagulation \revII{and fragmentation}. For the first time, we evolve a series of 2D simulation with dust coagulation over \SI{45000} planetary orbits and track the dust's size evolution and origin by using the \tpop{} dust coagulation method. We investigate the effects of different planetary masses, fragmentation velocities, and viscosities on the inner disk's dust mass budget and composition, and highlight the advantages of multi-dimensional simulations over 1D models. Filtering can only be efficient for high planetary masses, high fragmentation velocities, and low diffusivities. Clear compositional distinctions between the inner and outer disk could not have been maintained by Jupiter's core if the fragmentation velocity was low, even if $\alpha \lesssim 5\times 10^{-4}$. Significant ``contamination'' of the inner disk by outer-disk dust occurs in much less than \SI{2e5}{\years} in this case and even for more massive objects. This either places tight constraints on the physical conditions in the Solar nebula or mandates consideration of alternative explanations for the NC-CC dichotomy. Astrophysical constraints on the parameters could discriminate between these possibilities.

\end{abstract}


\section{Introduction} \label{sec:intro}
Gas giant planets form through runaway gas accretion onto a rocky core \citep{Mizuno1978,Pollack1996} or via gravitational instability of the protoplanetary disk \citep{Boss1997}.
While conditions for gravitational instability \citep{Kratter2016} may be met in the cool outer parts of protoplanetary disks \citep{Clarke2009,Rafikov2009}, core accretion is the probable origin of giant planets in the inner disk \citep[see][and references therein]{Drazkowska2022}. 
Once a critical core mass is reached, runaway gas accretion sets in forming a massive, gravitationally bound, gaseous envelope around the core. 
The protoplanet's gravitational field then exerts perturbations on the surrounding gas disk, creating a largely axisymmetric annular gap. Exterior to the gap, a locally positive radial pressure gradient means that the gas orbits at super-Keplerian speed, rather than the sub-Keplerian rotation that is typical elsewhere. Streamers allow some gas to accrete onto the planet or pass through the gap region. \rev{These phenomena}, which have long been predicted \citep[e.g.][]{Lin1986,Lubow1999}, are now frequently observed in the form of substructures in protoplanetary disks \citep[e.g.,][]{Andrews2018}.

Any deviation of the gas from Keplerian rotation has a profound impact on the dynamics of the dust component. 
Due to the negligible impact of pressure gradients on grains, their dynamics are distinct from the gas---yet coupled via aerodynamic drag forces that lead to momentum exchange with the gas. Dust particles make up a small fraction of the total disk mass (nominally \SI{1}{\percent}) and thus mostly follow the typically sub-Keplerian orbits of the gas. Lacking the gas' radial pressure support, they slowly fall onto the central star in a process called radial drift \citep{Whipple1972, Weidenschilling1977}. At the gap edge, however, this radial drift can be reversed and completely stopped at the location of a local pressure maximum \citep{Whipple1972}. To leading order, planet-induced gaps thus act as barriers to inward drifting dust particles, with various implications for the compositional evolution of protoplanetary disks. For example, water and volatile rich dust from the outer disk may not be able to reach the inner disk \citep{Kalyaan2023}. The possible presence of drift barriers is particularly interesting in the Solar System context, where the meteoritic record shows that planetesimal formation must have occurred in two spatially and temporally separated reservoirs of mass with distinct nucleosynthetic composition. Measured against terrestrial isotopic standards, chondrites fall into two broad categories known as carbonaceous and non-carbonaceous chondrites \citep[][and references therein]{Trinquier2007, Warren2011, Kleine2020}. The formation of Jupiter is often evoked as an explanation for why these two populations have been separated for extended amounts of time \citep[e.g.,][]{Kruijer2017, Morbidelli2024}.

The ability of proto-Jupiter to maintain a compositional dichotomy in the Solar Nebula for an extended period of time depends on the dust size distribution.
While large grains can be efficiently trapped at the outer edges of gaps carved by planets, smaller particles tend to closely follow the gas' motion and diffuse through the trap---a process referred to as \textit{dust filtration} \citep{Paardekooper2006,Rice2006,Ward2009}. In simulations including dust diffusion, \cite{Zhu2012} found that significant amounts of small grains could reach the inner disk for $\alpha=10^{-2}$, in part also due to the high gas accretion fluxes. The works by \cite{Weber2018} showed that only particles above a certain Stokes number (or aerodynamic size) can be effectively stopped from moving inwards for a given diffusivity.
Three-dimensional studies raised further concerns about a gap's filtering efficiency, as small particles can be lifted to larger altitudes and pass the growing planet on more meridional trajectories \citep{Binkert2021, Huang2025}. Dust flows across gaps are consistent with some astronomical constraints, for example in the PDS-70 system where even multiple massive planets are not acting as perfect dust traps, letting some amount of dust leak into the inner cavity \citep{Sierra2025}. In the Solar System, however, the presence of small grains with the ability to pass through Jupiter's co-orbital region could dilute or completely lift the observed isotopic dichotomy---making alternative explanations necessary \citep[\rev{see, e.g.,}][]{Brasser2020}.

Tracking the size distribution of the dust throughout the evolution of a protoplanetary disk is thus key to quantify the filtering efficiency of planetary gaps. Dust grains undergo continuous collisions, leading to the fragmentation (break-up) and coagulation (sticking) of the interacting particles, depending on the material properties and collision velocities \citep[see][and references therein]{Birnstiel2025}.
The amount of small grains that can traverse the gap by following the gas flows is thus crucially dependent on the grain-grain collision physics. 
Various studies have investigated the influence of coagulation and fragmentation on a gap's ability to stop inwards migrating grains.
\cite{Drazkowska2019} investigated the impact of dust coagulation in two-dimensional hydrodynamic simulations of protoplanetary disks with the code \LACOMPASS{}. They showed significant leakage of small grains can occur if fragmentation operates at the outer gap edge.
This study was, however, limited in scope and runtime by the high computational cost of the coagulation model.
One dimensional studies are, thus, oftentimes used to study the impact of substructures on the dust dynamics. 
\cite{Stammler2023} investigated the filtering efficiency of a giant planet core under various scenarios and concluded that it could not pose a significant hurdle for the inwards drifting and diffusing dust particles, produced via fragmentation at the gap edge.
\cite{Haugbolle2019} and \cite{Homma2024} refined such one-dimensional studies by explicitly tracking the grain composition throughout the disk. \cite{Homma2024} in particular found that isotopic variations in CI chondrites in the inner disk can be plausibly explained by $^{54}\mathrm{Cr}$-rich particles that diffuse through the gap.
One-dimensional models, however, cannot account for non-axisymmetric flows around a planet and the effects of spiral density waves, which might also influence the particle sizes throughout the disk \citep{Drazkowska2019, Eriksson2025}.



We aim to investigate under what circumstances a planet-induced gap can maintain a given compositional dichotomy in a planet-disk system by use of hydrodynamic simulations including dust coagulation. The simplified coagulation model \tpop{} makes it possible to study the evolution of the local grain size distribution at relatively low computational cost and therefore allows us to track the dynamics of small and large grains and the material flux through the disk.
Since two-dimensional simulations with \tpop{} are inexpensive to run, we probe a small parameter space of different diffusivities, fragmentation velocities and planetary masses---all of which are super-thermal.

We introduce \tpop{} \citep{Pfeil2024a} and the numerical methods we implemented in the \athena{} multifluid MHD code \citep{Stone2020, Huang2022} for this work in \autoref{sec:Methods}.
Our results \autoref{sec:results} is structured as follows: We present a general overview of the global disk evolution for different planetary masses, dust fragmentation velocities, and viscosities/dust diffusivies in \autoref{sec:Mplanet}, \autoref{sec:vfrag}, \autoref{sec:alpha}, respectively.
In \autoref{sec:CompositionalEvolution}, we investigate the respective compositional evolution in the inner disk due to influx of outer-disk material in the respective simulations. \autoref{sec:FringeCase} focuses on the special case of a low-viscosity, low-fragmentation-velocity disk with a planet that has approximately the mass of Jupiter's core.
We then discuss and sum up our results in \autoref{sec:discussion} and \autoref{sec:summary} respectively.

\section{Numerical Methods}\label{sec:Methods}
\subsection{\tpop{} Dust Coagulation Model}
\tpop{} \citep{Pfeil2024b} evolves a polydisperse dust size distribution under the assumption that the dust densities can be expressed as truncated power laws of the particle size (typically defined via the number density distribution $\Diff{n(a)}{a}\propto a^q$), between a minimum and maximum particle size ($\amin$ and $\amax$ respectively). 
The dust \rev{column} density distribution can then be written
\begin{align}
   \Diff{\Sigma_\mathrm{d}(a)}{a}= \begin{cases}
    \dfrac{\Sigma_{\text{d,tot}}(q+4)}{\amax^{q+4} - \amin^{q+4}} a^{q+3} & \text{for } q\neq -4 \\[15pt]
    \dfrac{\Sigma_{\text{d,tot}}}{\log(\amax/\amin)}\dfrac{1}{a} & \text{for } q=-4\, ,
  \end{cases}
\end{align}
where $\Sigma_\mathrm{d,tot}=\int \Diff{\Sigma_\mathrm{d}(a)}{a}\mathrm{d}a$ is the total dust density.
In \tpop{}, these size distributions are split into two bins, separated by an intermediate size $\aint=\sqrt{\amin\amax}$
\begin{align}
\begin{split}
     \rhodsm &= \int_{a_{\mathrm{min}}}^{\aint} \Diff{\Sigma_\mathrm{d}(a)}{a}\,\mathrm{d}a 
     \\ 
     \rhodlr &= \int_{\aint}^{a_{\mathrm{max}}} \Diff{\Sigma_\mathrm{d}(a)}{a}\,\mathrm{d}a\, 
\end{split}
\end{align}
which means the size distribution exponent is defined as
\begin{equation}
    q=2\frac{\log\left(\rhodlr/\rhodsm\right)}{\log\left(\amax/\amin\right)} - 4\, .
\end{equation}
This makes it possible to capture the dust-dust interactions by evolving only two dust fluids and the maximum grain size $\amax$ instead of numerically solving the coagulation equation \citep{Smoluchowski1916}.

Coagulation and fragmentation are modeled as collisional mass exchange rates between the two populations (``coagulation'': $\dot{\Sigma}_{\mathrm{d},0\rightarrow 1}$; ``fragmentation'': $\dot{\Sigma}_{\mathrm{d},1\rightarrow 0}$), which are calibrated to reproduce the results of established Smoluchowski solvers such as \dpy{} \citep{Stammler2022}.
\rev{To accurately reproduce the dust flux and the maximum particle sizes within planetary gaps, an additional source term \revII{for $\amax$ and $\Sigma_{\mathrm{d},1}$} was introduced \citep[see][section 4.3]{Pfeil2024b}. This term ensures that the size distribution in the low density regime of planet-induced gaps, where the collision rates are very low, retains a meaningful exponent $q$ and cutoff size $a_\mathrm{max}$.}

Given the local values of $\amax$ and $q$, we can calculate the mass-averaged particle size of a size interval $[a_\RN{1},a_\RN{2}]$
\begin{align}
\begin{split}
    \langle a\rangle_{a_\RN{1}}^{a_\RN{2}} &=  \dfrac{\int_{a_\RN{1}}^{a_\RN{2}} \Diff{\Sigma_\mathrm{d}(a)}{a}\,a\,\mathrm{d}a}{\int_{a_\RN{1}}^{a_\RN{2}} \Diff{\Sigma_\mathrm{d}(a)}{a}\, \mathrm{d}a}  \\
&= \begin{cases}
    \dfrac{a_\RN{2}a_\RN{1}}{a_\RN{2}-a_\RN{1}}\log\left(\dfrac{a_\RN{2}}{a_\RN{1}}\right) & \text{for } q=-5 \\[15pt]
    \dfrac{q+4}{q+5} \dfrac{a_\RN{2}^{q+5} - a_\RN{1}^{q+5}}{a_\RN{2}^{q+4} - a_\RN{1}^{q+4}} & \text{for } q\neq -5,-4 \\[15pt]
    \dfrac{a_\RN{2}-a_\RN{1}}{\log(a_\RN{2})-\log(a_\RN{1})} & \text{for } q=-4\, ,
  \end{cases} \label{eq:amean}
\end{split}
\end{align}
which defines our two representative particle sizes $a_0\equiv\langle a\rangle_{\amin}^{\aint}$ and $a_1\equiv\langle a\rangle_{\aint}^{\amax}$.

On a discrete grid of sizes, the individual density values for each bin are given by
\begin{equation}
    \Sigma_{\mathrm{d},i}=\int_{a_{i-\nicefrac{1}{2}}}^{a_{i+\nicefrac{1}{2}}} \Diff{\Sigma_\mathrm{d}(a)}{a}\mathrm{d}a\, ,
\end{equation}
and thus depend on the grid spacing. When plotting the complete size distribution we thus show the density per logarithmic size bin
\begin{align}
    \sigma_\mathrm{d} =\Diff{\Sigma_\mathrm{d}(a)}{\log(a)} =  \begin{cases}
    \dfrac{\Sigma_{\text{d,tot}}(q+4)}{\amax^{q+4} - \amin^{q+4}} a^{q+4} & \text{for } q\neq -4 \\[15pt]
    \dfrac{\Sigma_{\text{d,tot}}}{\log(\amax/\amin)} & \text{for } q=-4\, ,
  \end{cases}
\end{align}
which is independent of the grid spacing. 

\subsection[Implementation in Athena++]{Implementation in \athena{}}
We employ the \athena{} multifluid module \citep{Huang2022} to model dust dynamics. Since we set up our simulations in terms of the vertically integrated disk structure, we use column densities $\Sigma$ instead of volume densities $\rho$. 
The conservative equations for the gas remain unaltered in our simulations
\begin{align}
\diff{\rhog}{t} + \boldsymbol{\nabla} \cdot (\rhog \vg) = 0 \, , \label{eq:CEq}
\end{align}
\begin{align}
&\diff{}{t} (\rhog \vg) + \boldsymbol{\nabla} \cdot (\rhog \vg \vg + P_{\mathrm{g}} \mathbb{I} + \boldsymbol{\Pi}_{\nu}) \nonumber \\
&= \rhog \vect{f}_{\mathrm{g,src}} + \sum_{n=1}^{N_{\mathrm{d}}} \rhodn \frac{\vdn - \vg}{T_{\mathrm{s},n}}\, , \label{eq:NSEq} 
\end{align}
Subscripts ``g'' and ``d'' refer to the gas and dust quantities respectively. Velocity vectors are given as $\vect{v}$, gas pressure is denoted as $P_\mathrm{g}$, $\boldsymbol{\Pi}_\nu$ is the viscous stress tensor, and additional source terms are given as $\vect{f}_\mathrm{src}$. We use the \cite{Shakura1973} viscosity model, with $\nu=\alpha c_\mathrm{s} H$, where $\alpha$ is the viscosity parameter, $c_\mathrm{s}$ is the soundspeed, and $H$ is the gas pressure scale height.
Dust-gas feedback for $N_\mathrm{d}$ dust species is incorporated via the drag force term in \autoref{eq:NSEq}. The stopping time $T_{\mathrm{s},n}$ determines the strength of the dust-gas coupling and is defined in \autoref{sec:MaxPartSizeEvol}.
In the following simulations we enforce a locally isothermal EOS, which is why we omit the energy equation here.

To model dust coagulation with \tpop{} in the \athena{} multifluid module, the continuity and momentum equations of two dust fluids ($N_\mathrm{d}=2$), representing $\rhodsm$ and $\rhodlr$, must be coupled to account for the dust-dust interactions. This is realized via mass exchange terms $\dot{\Sigma}_{\mathrm{d},n}$ \citep[see Equation 35 in][]{Pfeil2024b} and momentum exchange terms $\boldsymbol{\dot{\mathcal{M}}}_{\mathrm{d},n}$
\begin{align}
&\diff{\rhodn}{t} + \boldsymbol{\nabla} \cdot (\rhodn \vdn + \boldsymbol{\mathcal{F}}_{\mathrm{dif},n}) = \dot{\Sigma}_{\mathrm{d},n} \, , 
\\ \nonumber
\\
&\diff{}{t} (\rhodn (\vdn + \vect{v}_{\mathrm{d,dif},n})) 
+ \boldsymbol{\nabla} \cdot (\rhodn \vdn \vdn) \nonumber 
\\
&= \rhodn \vect{f}_{\mathrm{d,src},n} + \rhodn \frac{\vg - \vdn}{T_{\mathrm{s},n}} + \boldsymbol{\dot{\mathcal{M}}}_{\mathrm{d},n}\, . \label{eq:NSEqD}
\end{align}
The mass transfer from the small population to the large population ($\dot{\Sigma}_{\mathrm{d},0\rightarrow 1}$) and from the large population to the small population ($\dot{\Sigma}_{\mathrm{d},1\rightarrow 0}$) are given in \cite{Pfeil2024b} and define the source terms as 
\begin{align}
    \dot{\Sigma}_\mathrm{d,0} &= \dot{\Sigma}_{\mathrm{d},1\rightarrow 0} - \dot{\Sigma}_{\mathrm{d},0\rightarrow 1}  \\ 
    \dot{\Sigma}_\mathrm{d,1} &= \dot{\Sigma}_{\mathrm{d},0\rightarrow 1} - \dot{\Sigma}_{\mathrm{d},1\rightarrow 0} \, ,
\end{align}
thus, conserving the total dust mass.
Coagulating/fragmenting particles are assumed to carry their momenta between the two populations. The momentum transfer rates are calculated accordingly as
\begin{align}
    \boldsymbol{\dot{\mathcal{M}}}_{\mathrm{d},0\rightarrow 1} &= \dot{\Sigma}_{\mathrm{d},0\rightarrow 1} \vect{v}_{\mathrm{d},0} \\
    \boldsymbol{\dot{\mathcal{M}}}_{\mathrm{d},1\rightarrow 0} &= \dot{\Sigma}_{\mathrm{d},1\rightarrow 0} \vect{v}_{\mathrm{d},1}
\end{align}
and thus, under conservation of momentum, we have
\begin{align}
    \boldsymbol{\dot{\mathcal{M}}}_{\mathrm{d},0} &= \boldsymbol{\dot{\mathcal{M}}}_{\mathrm{d},1\rightarrow 0} - \boldsymbol{\dot{\mathcal{M}}}_{\mathrm{d},0\rightarrow 1}\\
    \boldsymbol{\dot{\mathcal{M}}}_{\mathrm{d},1} &= \boldsymbol{\dot{\mathcal{M}}}_{\mathrm{d},0\rightarrow 1} - \boldsymbol{\dot{\mathcal{M}}}_{\mathrm{d},1\rightarrow 0}\, .
\end{align}
The dust fluids undergo concentration diffusion based on the local gradient in the dust-to-gas ratios
\begin{equation}
       \boldsymbol{\mathcal{F}}_\mathrm{dif,n} = - \rhog 
    D_n\boldsymbol{\nabla}\left(\frac{\rhodn}{\rhog}\right)\rev{ = \Sigma_\mathrm{d,n}\vect{v}_{\mathrm{d,dif},n}}\,,
\end{equation}
\rev{with the dust diffusion velocity $\vect{v}_{\mathrm{d,dif},n}$}. Here, $D_n$ is the dust diffusivity, defined as 
\begin{equation}
    D_n = \frac{\alpha c_\mathrm{s} H}{1+\mathrm{St}_n^2}\,,
\end{equation}
where $\mathrm{St}_n$ is the respective dust species' representative Stokes number.
We only include the diffusion term in the continuity equation for now and leave studies including momentum diffusion for future investigations.

\subsubsection{Maximum Particle Size Evolution and Stopping Times} \label{sec:MaxPartSizeEvol}
The maximum particle size is evolved as a passive scalar which follows the large dust fluid ($n=1$) and evolves based on the local fragmentation and coagulation process ($\dot{a}_\mathrm{max}$)
\begin{align}
&\diff{\rhodlr \amax}{t} + \boldsymbol{\nabla} \cdot (\amax\rhodlr \vect{v}_{\mathrm{d},1}\revII{+\amax\boldsymbol{\mathcal{F}}_{\mathrm{dif,1}}}) \\&=\boldsymbol{\nabla} \cdot \boldsymbol{\mathcal{F}}_{\mathrm{dif,}\amax}\nonumber
+ \rhodlr \dot{a}_\mathrm{max} + \amax \dot{\Sigma}_\mathrm{d,1}\, . 
\end{align}
The maximum particle size is diffused based on the local gradient in particle size
\begin{equation}
        \boldsymbol{\mathcal{F}}_{\mathrm{dif},\amax} = - \rhodlr D_1\boldsymbol{\nabla}\amax \, .
\end{equation}
For details on the evolution of $\amax$, we refer to Equation 29 in \cite{Pfeil2024b}.
Given the local values of $\amax$ and $\aint$, we calculate the mass-averaged particle sizes $a_0$ and $a_1$ in every grid cell via \autoref{eq:amean}. 
These define the dust-gas stopping times $T_{\mathrm{s},n}$ in the Epstein regime, 
\begin{equation}
    T_{\mathrm{s},n} = \frac{\pi}{2}\frac{a_{n}\rho_\mathrm{m}}{\rhog \Omega_\mathrm{K}}\, ,
\end{equation}
which are used to determine the dust-gas drag forces in \autoref{eq:NSEq} and \autoref{eq:NSEqD}.

\subsubsection{Tracking Compositional Evolution due to Transport and Coagulation/Fragmentation}\label{sec:PassiveScalars}
In order to track the compositional changes of the dust due to leakage through a planetary gap, we introduce two passive scalars, which follow our two dust fluids:
\begin{align}
    &\diff{\rhodsm C_0}{t} + \boldsymbol{\nabla} \cdot (C_0\rhodsm \vect{v}_{\mathrm{d},0}) &= \boldsymbol{\nabla} \cdot \boldsymbol{\mathcal{{F}}}_{\mathrm{dif},C_0} + \mathcal{\dot{C}}_0 \\
    &\diff{\rhodlr C_1}{t} + \boldsymbol{\nabla} \cdot (C_1\rhodlr \vect{v}_{\mathrm{d},1}) &= \boldsymbol{\nabla} \cdot \boldsymbol{\mathcal{{F}}}_{\mathrm{dif},C_1} +  \mathcal{\dot{C}}_1\end{align}
$\mathcal{\dot{C}}_{0/1}$ denotes the mass exchange between the two scalars based on the coagulation/fragmentation rates. Composition is carried by the fragmenting/coagulating particles in the same way as momentum.  
\begin{align}
\mathcal{\dot{C}}_{0\rightarrow 1} &=  \dot{\Sigma}_{\mathrm{d},0\rightarrow 1} C_0 \\
\mathcal{\dot{C}}_{1\rightarrow 0} &=  \dot{\Sigma}_{\mathrm{d},1\rightarrow 0} C_1.
\end{align}
Under conservation of mass, we can thus write
\begin{align}
    \dot{\mathcal{C}}_0 &= \mathcal{\dot{C}}_{1\rightarrow 0} - \mathcal{\dot{C}}_{0\rightarrow 1} \\
     \dot{\mathcal{C}}_1 &= \mathcal{\dot{C}}_{0\rightarrow 1} - \mathcal{\dot{C}}_{1\rightarrow 0}\, .
\end{align}
The passive scalars also undergo gradient diffusion, similar to the dust species and the maximum particle size
\begin{align}
    \boldsymbol{\mathcal{F}}_{\mathrm{dif},C_n} = - \Sigma_{\mathrm{d},n} D_n\boldsymbol{\nabla}C_n \, , 
\end{align}
where the diffusivity $D_n$ is equal to the dust diffusivity.

\subsection[Adjustments with respect to T.\ Pfeil et al.\ (2024)]{Adjustments with respect to \cite{Pfeil2024b}}
In comparison to the original version of \tpop{} \citep{Pfeil2024b} in the \PLUTO{} code \citep{Mignone2007}, some differences exist to the \athena{}-\tpop{} version:
\begin{itemize}
    \item In \PLUTO{}, the two dust fluids were implemented as passive scalars which were given the dust-gas terminal velocity \citep{Nakagawa1986} to simulate the effects of gas drag. These velocities were calculated based on the disk's midplane pressure gradient to allow for a detailed comparison with \dpy{}. In our vertically-integrated \athena{} simulations, the sub-Keplerian motion of the gas is the result of the vertically integrated pressure gradient. The dust-gas drag forces in our \athena{} simulations thus result in slightly different drift velocities than in \PLUTO{}.
    \item Due to the terminal velocity assumption in \PLUTO{}, the Stokes numbers were limited to $<1$. In the \athena{} multi-fluid module, Stokes number can take arbitrary values. The fluid approximation, however, is only valid for $\St<1$. This is only relevant for models with very large particles or in severely gas-depleted regions.  
    \item Since dust fluids in \PLUTO{} were passive scalars, no momentum equation had to be solved and no momentum exchange between the populations had to be considered.
    Dust and gas in the \athena{} version are dynamically coupled through the drag forces.
\end{itemize}

\subsection{Grid and initial conditions}
We set up simulations on a logarithmically spaced radial grid, spanning a range from \SI{1.56}{\AU} to \SI{20.8}{\AU} with 600 grid cells. 
We resolve the full azimuthal dimension at a resolution of 528 grid cells.
Our initial density profile follows the \cite{LyndenBell1974} profile with a power-law exponent $\beta_\Sigma=-1$ and a characteristic disk radius of $R_\mathrm{char}=\SI{60}{\AU}$. The disk temperature is given by
\begin{equation}
    T = \left(\frac{\Theta}{2} \frac{L_*}{4\pi R^2 \sigma_\mathrm{SB}}\right)^{\nicefrac{1}{4}},
\end{equation}
where $L_*$ is the stellar luminosity, $R$ is the cylindrical stellocentric distance, and $\sigma_\mathrm{SB}$ is the Stefan-Boltzmann constant. $\Theta$ represents the grazing angle of the disk's photosphere which is set to $\Theta=0.02$ \citep[adopted from][]{Dullemond2018} throughout all simulations.
We adopt the first tabulated luminosity value from the \cite{Baraffe2015} pre-main-sequence evolutionary track of a \SI{1}{\Msol} star \rev{(constant $L_*=\SI{1.24e34}{\erg\per\second}=\SI{3.22}{\Lsol}$). This leads to a temperature of \SI{73.2}{\kelvin} at \SI{5.2}{\AU}. Passively irradiated disk's like we consider them here, are generally cooler than viscously heated disks.}

\begin{figure}[ht]
    \centering
    \includegraphics[width=\linewidth]{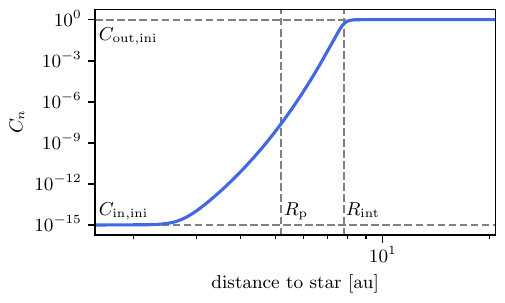}
    \caption{Initial distribution of the passive scalar for both dust fluids. This quantity represents the initial distribution of outer-disk material in our simulations.}
    \label{fig:InitialConc}
\end{figure}

In order to track the amount of material that diffuses through the planetary gap, we initialize the two passive scalars (see \autoref{sec:PassiveScalars}) as smooth, step-like functions
\begin{equation}
    \Sigma_{\mathrm{d,out},n} = \Sigma_{\mathrm{d},n}  \left(C_\mathrm{in,ini} + \frac{C_\mathrm{out,ini}-C_\mathrm{in,ini}}{1 + e^{-35(R-R_\mathrm{int})}}\right),
\end{equation}
with $R_\mathrm{int}=1.5R_\mathrm{p}=\SI{7.8}{\AU}$, where $R_\mathrm{p}=\SI{5.2}{\AU}$ is the planet's orbital distance to the star in all simulations (see \autoref{fig:InitialConc}).
The value of our passive scalar thus represents the local concentration of outer-disk dust.

We ensure a constant inflow of dust throughout the simulation by setting the total dust-to-gas ratio in the outer ghost zones to the initial value (\SI{1}{\percent}).
\begin{figure*}[ht]
    \centering
    \includegraphics[width=\linewidth]{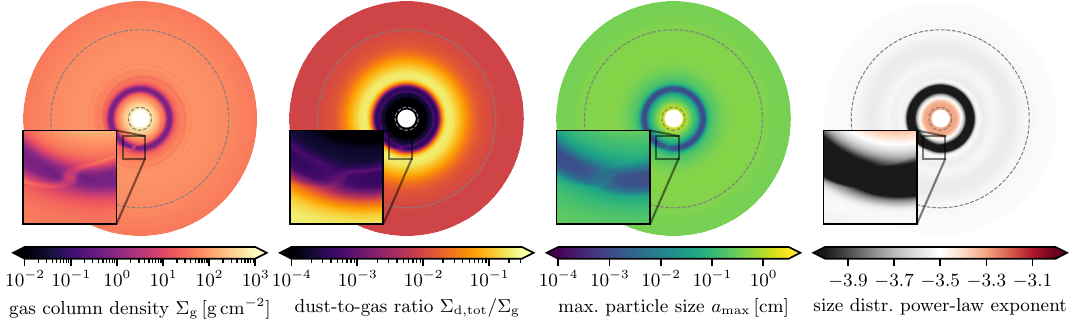}
    \caption{Snapshot of our fiducial simulation with $M_\mathrm{p}=\SI{0.4}{\Mjup}$, $v_\mathrm{frag}=\SI{400}{\centi\meter\per\second}$ and $\alpha=10^{-3}$ after \SI{45000}{} planetary orbits. We employ damping zones in the boundary layers, marked my the thin dashed lines.}
    \label{fig:Fiducial}
\end{figure*}

\subsection{Planetary gravitational potential and wave-damping zones}

We include the gravitational potential of a planet of mass $M_\mathrm{p}$, located at a distance $R_\mathrm{p}$ from the central star and at azimuth $\varphi_\mathrm{p}$ in our simulations. At a distance $d^2=R^2+R_\mathrm{p}^2 -2RR_\mathrm{p}\cos(\varphi-\varphi_\mathrm{p})$ to the planet we prescribe the planet's gravitational effect as a Plummer potential
\begin{equation}
    \Phi_\mathrm{p}^* = -\frac{\mathrm{G}M_\mathrm{p}}{\sqrt{d^2 + r_\mathrm{sm}^2}}\, ,
\end{equation}
where $\mathrm{G}$ is the gravitational constant and $r_\mathrm{sm}=\Psi a_\mathrm{H}$ is called the smoothing length. We define it in terms of the planet's Hill radius
\begin{equation}
    a_\mathrm{H}=R_\mathrm{p}\left(\frac{1}{3}\frac{M_\mathrm{p}}{M_\mathrm{p}+M_*}\right)^{\nicefrac{1}{3}}
\end{equation} with $\Psi=0.7$. \rev{This value is similar the ones used by \cite{Weber2018} ($r_\mathrm{sm}=0.6 H$), and \cite{Huang2025} ($\Psi=0.6$).}
To account for the acceleration of the star in our stellocentric coordinate system, we add an indirect potential \citep[see e.g.][]{Weber2018}
\begin{align}
    \Phi_\mathrm{ind}=\frac{GM_\mathrm{p}}{R_\mathrm{p}^2}R\cos(\varphi-\varphi_\mathrm{p})
\end{align}
to $\Phi_\mathrm{p}^*$.
The respective gravitational accelerations are implemented as source terms in \autoref{eq:NSEq} and \autoref{eq:NSEqD} 
\begin{align}
    \vect{f}_\mathrm{g,src} &= - \rhog\left(\mathrm{G}M_\mathrm{p}\frac{\boldsymbol{\nabla}d^2}{2d^3} + \boldsymbol{\nabla}\Phi_\mathrm{ind}.\right) \\
    \vect{f}_{\mathrm{d,src},n} &= - \Sigma_{\mathrm{d},n}\left(\mathrm{G}M_\mathrm{p}\frac{\boldsymbol{\nabla}d^2}{2d^3} + \boldsymbol{\nabla}\Phi_\mathrm{ind}\right).
\end{align}

We apply wave-damping layers, as in \cite{deValBorro2006} to the outer and inner radial boundary regions of our simulations. The extend of these layers is defined by the orbital period ratio $\Xi=\Omega_\mathrm{K,in,b}/\Omega_\mathrm{K,in}=\Omega_\mathrm{K,out}/\Omega_\mathrm{K,out,b}$ between the inner/outer domain edges $R_\mathrm{in}$ and $R_\mathrm{out}$ and the interior edges of the damping layers $R_\mathrm{in,b}$ and $R_\mathrm{out,b}$
\begin{align}
    \Delta R_\mathrm{damp, in} &= R_\mathrm{in}(\Xi^{\nicefrac{2}{3}}-1) \\
    \Delta R_\mathrm{damp, out} &= R_\mathrm{out}(1-\Xi^{-\nicefrac{2}{3}})\, .
\end{align}
We adjust the respective conservative variables over the timestep $\Delta t$ by an amount
\begin{equation}
    \Delta X = \left[\exp\left(-\frac{\Delta t\, f}{\tau_\mathrm{damp}}\right)-1\right](X-X_0)\, , \label{eq:damp}
\end{equation}
such that $X^{n+1} = \rev{X^{n}}+\Delta X$. The quantities are relaxed towards a defined equilibrium state $X_0$ on a timescale $\tau_\mathrm{damp}=\chi\Omega_K^{-1}$. The function $f$ quadratically increases the damping efficiency from zero to unity towards the boundary in order to avoid unsteady behavior at the boundary layer edges:
\begin{equation}
f = \left[\frac{\max(0,R_\mathrm{in,b}-R)}{\Delta R_\mathrm{damp, in}} + \frac{\max(0,R-R_\mathrm{out,b})}{\Delta R_\mathrm{damp, out}}\right]^2
\end{equation}
Using \autoref{eq:damp} instead of the explicit damping term prevents numerical overshooting and makes the method stable for arbitrary time steps.
In our simulations we use $\Xi=1.5$ and $\chi=0.3$ if not noted otherwise.

\begin{table}[ht]
    \caption{Parameters and results of our simulations. The left four columns show the input parameters for our simulations\rev{, where $M_\mathrm{th}$ denotes the thermal mass}. The fifth and sixth columns shows the average concentration and the total mass of dust in the inner disk that originates from the outer disk at the end of the simulations (after \SI{45000}{} orbits).}
    \centering
    \begin{tabular}{cccc|cc}
    \toprule
    \multicolumn{2}{c}{$M_\mathrm{p}\, $} & $v_\mathrm{frag}$ & $\alpha$ & $\langle C_\mathrm{out}\rangle_\mathrm{in}$ & $M_{\mathrm{d,out\rightarrow in}}$ \\
    $[M_\mathrm{Jup}]$ & $[M_\mathrm{th}]$ & $[\mathrm{cm \, s^{-1}}]$ & & [\%] & $[M_{\oplus}]$ \\
    \midrule \midrule
{\bfseries 0.4} & {\bfseries 6.4} & {\bfseries 400} & $\boldsymbol{10^{-3}}$ & {\bfseries 84.0} & {\bfseries 0.0236} \\
    \midrule
    0.1 & 1.6  & 50 & $2.5\times 10^{-4}$ & 21.5 & 0.4441 \\
    \midrule
    0.1 & 1.6  & 400 & $10^{-3}$ & 82.5 & 1.9618\\
    0.2 & 3.2  & 400 & $10^{-3}$ & 82.7 &  0.2589\\
    0.6 & 9.7  & 400 & $10^{-3}$ & 84.0 & 0.0051\\
    \midrule
    0.4 & 6.4 & 100 & $10^{-3}$ & 62.1 & 1.7324 \\ 
    0.4 & 6.4 & 200 & $10^{-3}$ & 73.1 & 0.5241 \\ 
    0.4 & 6.4 & 800 & $10^{-3}$ & 88.7 & 0.0049 \\ 
    \midrule
    0.4 & 6.4 & 400 & $5\times 10^{-4}$ & 62.7 & 0.0002 \\
    0.4 & 6.4 & 400 & $2\times 10^{-3}$ & 78.5 & 2.4397 \\
    0.4 & 6.4 & 400 & $4\times10^{-3}$ & 75.8 & 4.9492 \\
         \bottomrule
    \end{tabular}
    \label{tab:ParameterStudy}
\end{table}

\section{Results} \label{sec:results}
\begin{figure}[ht]
    \centering
    \includegraphics[width=\linewidth]{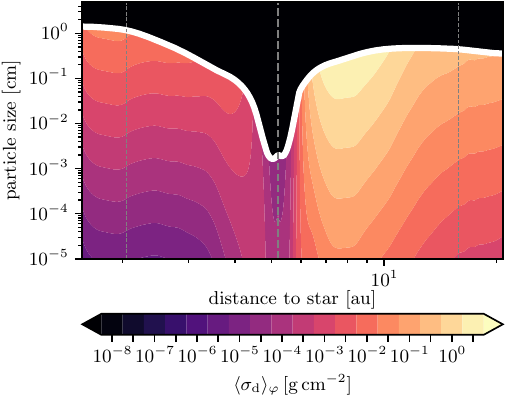}
    \caption{Azimuthally averaged dust size distribution of our fiducial simulation after \SI{45000}{} planetary orbits. The maximum particle size is shown in white. The planet is located at \SI{5.2}{\AU}. We employ damping zones in the boundary layers, marked \rev{by} the thin dashed lines.}
    \label{fig:FullDistrFid}
\end{figure}
We run our simulations for \SI{45000}{} planetary orbits ($\sim\SI{533602}{\years}$) to reach an equilibrium state. 
Even this extensive simulation runtime is not enough to reach a final equilibrium composition in the inner disk for some of the simulations. 
We chose a fragmentation velocity of $\SI{400}{\cm\per\second}$, an $\alpha$ parameter of $10^{-3}$, and a planetary mass of $\SI{0.4}{\Mjup} \approx \SI{6.4}{\Mth}$ as our nominal simulation setup. 
The last snapshot of the fiducial simulation is shown in \autoref{fig:Fiducial}.
The azimuthally averaged dust size distribution is depicted in \autoref{fig:FullDistrFid}. 

We show the nominal disk profile and the derived quantities in \Cref{fig:CompMp,fig:CompVfrag,fig:CompAlpha} as solid black lines.
As can be seen in the panel (a) of \autoref{fig:CompMp}, the planet in our nominal setup creates a deep gap in which the gas density is depleted by two orders of magnitude. 
Dust can be seen to accumulate at the outer gap edge, leading to an $\sim$ order of magnitude density enhancement in \autoref{fig:Fiducial} to \autoref{fig:CompMp}.
Particle drift is completely stopped at the gap edge.
In the nominal setup, particles can grow up to cm sizes outside of the planetary gap.
Larger particles are, however, efficiently removed from the planet's co-orbital region. The remaining dust particles have sizes of $\lesssim\SI{10}{\micron}$ and can only enter the gap via diffusion and by following the gas flow.
This constant dust flux is supported through the continuously replenished reservoir of fragmenting grains from the outer disk that collects in the pressure trap. The density of small particles is significantly enhanced within the pressure trap \rev{due to the trapping and fragmentation of large particles}, which can be seen in \autoref{fig:FullDistrFid} at $\sim\SI{8}{\AU}$. This reservoir of small dust is the main source for dust in the inner disk once the primordial grains have drifted out of the inner simulation domain.

Since fragmentation and coagulation maintain an MRN distribution at the gap edge, we can derive an analytic estimate for the fraction of small grains (i.e., $\rhodsm/\Sigma_\mathrm{d,tot}$) in the pressure trap.
Given a power-law exponent of $q=-3.5$ and a particle size of $\sim \SI{1}{\centi \meter}$ ($\chi\equiv\amax/\amin=10^5$), we find that the fraction of small grains in the pressure trap is approximately given by
\begin{equation}\label{eq:estimate}
    \frac{\rhodsm}{\Sigma_\mathrm{d,tot}} = \frac{1}{1+\chi^{\frac{q+4}{2}}} \approx 0.053\, ,
\end{equation}
which means \SI{5.3}{\percent} of the total dust mass in the pressure bump is present in the form of particles smaller than $\sqrt{\amax\amin}\approx\SI{31.6}{\micro \meter}$.
Continuous diffusion through the gap and drift of large particles into the pressure trap from the outer regions will thus provide a continuous diffusion flux of small particles through the gap.

Nonetheless, the ability of the gap edge to stop radial drift leads to significant depletion of the dust mass in the inner disk. Grains in the inner disk have grown to the fragmentation limit and drift faster inwards than diffusion can replenish them. 
We investigate the resulting compositional changes in more detail in \autoref{sec:CompositionalEvolution}.

\rev{
\subsection{Comparison to 1D models and the necessity for multi-dimensional simulations}
\begin{figure}[ht]
    \centering
    \includegraphics[width=\linewidth]{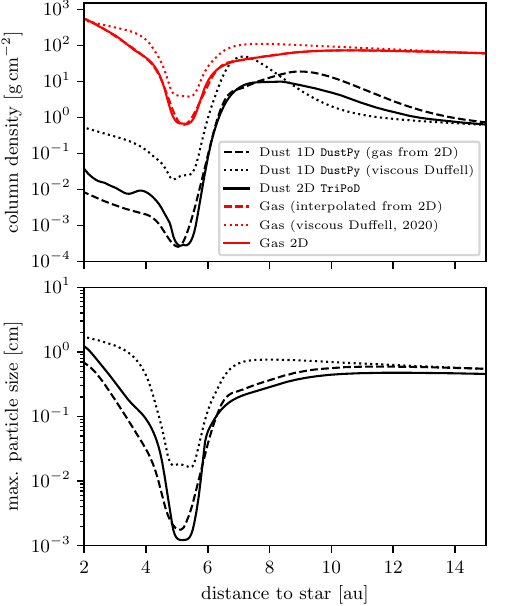}
    \caption{\rev{Comparison between the azimuthally averaged dust density and particle size profiles of our two-dimensional fiducial simulation (solid lines) and two one-dimensional models. Dashed lines show the one-dimensional \dpy{} model in which the gas column density was interpolated from the two-dimensional simulation results. Dotted lines correspond to a one-dimensional \dpy{} model for which the gas density was evolved based on a viscosity profile with in inverse \cite{Duffell2020} gap.}}
    \label{fig:1D_2D_Compare}
\end{figure}

\begin{figure}[ht]
    \centering
    \includegraphics[width=\linewidth]{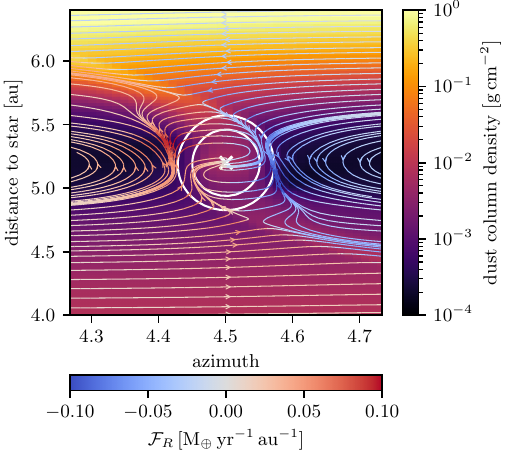}
    \caption{\rev{Streamlines of dust in the vicinity of the planet in our fiducial simulation. Azimuthal velocities are calculated in the planet's frame of reference. The color of the streamlines indicates the radial component of the flux. The white circles indicate the planet's Hill radius (outer circle) and the gravitational smoothing length (inner circle). Most of the mass flow between the inner and outer disk occurs via the flows past the planet.}}
    \label{fig:Streamlines}
\end{figure}
Capturing the geometry of the flows around the planet is of great importance for assessments of the dust filtration efficiency. 
\cite{Drazkowska2019} have shown the important differences between a two-dimensional and a one-dimensional simulation in detail. 

What exactly constitutes a ``fair'' comparison between one-dimensional and two-dimensional simulations is debatable. We have chosen a similar approach to \cite{Drazkowska2019} and \cite{Pinilla2024} and used the azimuthally averaged gas structure of our nominal two-dimensional simulation to inform a one-dimensional simulation with \dpy{} for comparison. 
For this, we employ the Clough-Tocher scheme, provided with the \texttt{SciPy} python package \citep{Virtanen2020}\footnote{\url{https://docs.scipy.org/doc//scipy-1.10.1/reference/generated/scipy.interpolate.CloughTocher2DInterpolator.html}} to interpolate the azimuthally averaged gas density profiles onto the \dpy{} grid and timestep. In this way, \dpy{} operates on the closest possible approximation to the gas structure from our two-dimensional model.
In order to achieve identical behavior at the boundaries, we furthermore introduce a damping zone in \dpy{} that relaxes the dust density profile towards the initial total dust density, while keeping the size distribution's shape unchanged. 
Any differences are thus most likely caused by the differences between one-dimensional and two-dimensional treatments and not by the methodological differences between \tpop{} and \dpy{} \citep[\tpop{} has been shown to accurately reproduce \dpy{} in ][]{Pfeil2024b}.
In accordance with \cite{Drazkowska2019}, we find that the dust densities in the inner disk are generally underestimated in the one-dimensional simulation compared to the two-dimensional \athena{} simulation (\autoref{fig:1D_2D_Compare}). 
By using an azimuthally averaged gas profile, we effectively average-out the gas and dust streams around the planet and thus diminish the radial flow of dust through the co-orbital region in our one-dimensional simulation. 
Furthermore, it can be seen that the flows past the planet contain slightly larger particles compared to the rest of the gap region (see \autoref{fig:Fiducial}). This means that a larger fraction of the entire dust size distribution can pass the gap at this location.
These effects have likely caused an underestimation of the dust leakage through the gaps in the one-dimensional simulations of PDS70 by \cite{Pinilla2024}. Larger particles than estimated by their models could have crossed the gaps in comparable two-dimensional simulations.

Most importantly, note that this close comparison is only possible by using the results of our two-dimensional simulation to inform the one-dimensional model.
Using the traditional one-dimensional treatment of planetary gaps results in a vastly different outcome. Instead of using the two-dimensional results, we also employ a \cite{Duffell2020} gap profile in the gas viscosity to create a gap in an alternative one-dimensional model.
Since this profile is significantly narrower and shallower than the gap we find in our two-dimensional simulation, orders of magnitude more dust can be found to make it into the inner disk in this case. The dust leakage in \citep{Stammler2023} might thus be overestimated in comparison to two-dimensional studies. 
The main advantage of two-dimensional and three-dimensional simulations is thus the self-consistent treatment of gap opening and the flow patterns around the planet, which cannot be captured by conventional one-dimensional approaches.

\autoref{fig:Streamlines} shows the detailed flow pattern around the planet. Since the gravitational smoothing radius covers most of the planet's Hill sphere, we cannot resolve the flows within a potential circumplanetary disk that might form in a higher resolution simulation with less gravitational smoothing.
It can be seen, however, that most of the dust flux reaching the inner disk only briefly enters the outer part of the planet's Hill sphere. 
\revII{Calculating the entire radial mass flux passing through the radial location of the planet, we can determine the fraction of the total radial mass flux within the Hill sphere.
At the end of the fiducial simulation approximately \SI{16}{\percent} of the total radial mass flux at the planet's radius flow through the Hill sphere. 
}
}

\subsection{Planetary Mass}\label{sec:Mplanet}
\begin{figure}[ht]
    \centering
    \includegraphics[width=\linewidth]{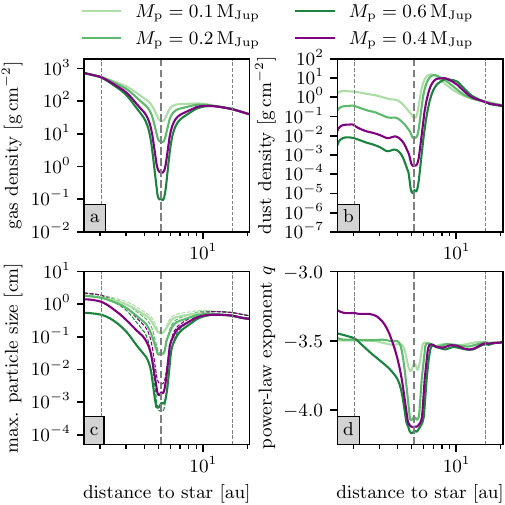}
    \caption{Azimuthally averaged disk profiles for the four simulations with different planetary masses after \SI{45000}{} orbits ($\sim\SI{533602}{\years}$) of evolution. Panel (a) shows the radial gas density profiles, panel (b) depicts the dust density profiles. Maximum particle sizes of the power-law size distributions are shown in panel (c). Thin dashed lines mark the analytically estimated fragmentation limit. The local power-law exponents of the dust size distributions are shown in panel (d).}
    \label{fig:CompMp}
\end{figure}

\begin{figure*}[!ht]
    \centering
    \includegraphics[width=\linewidth]{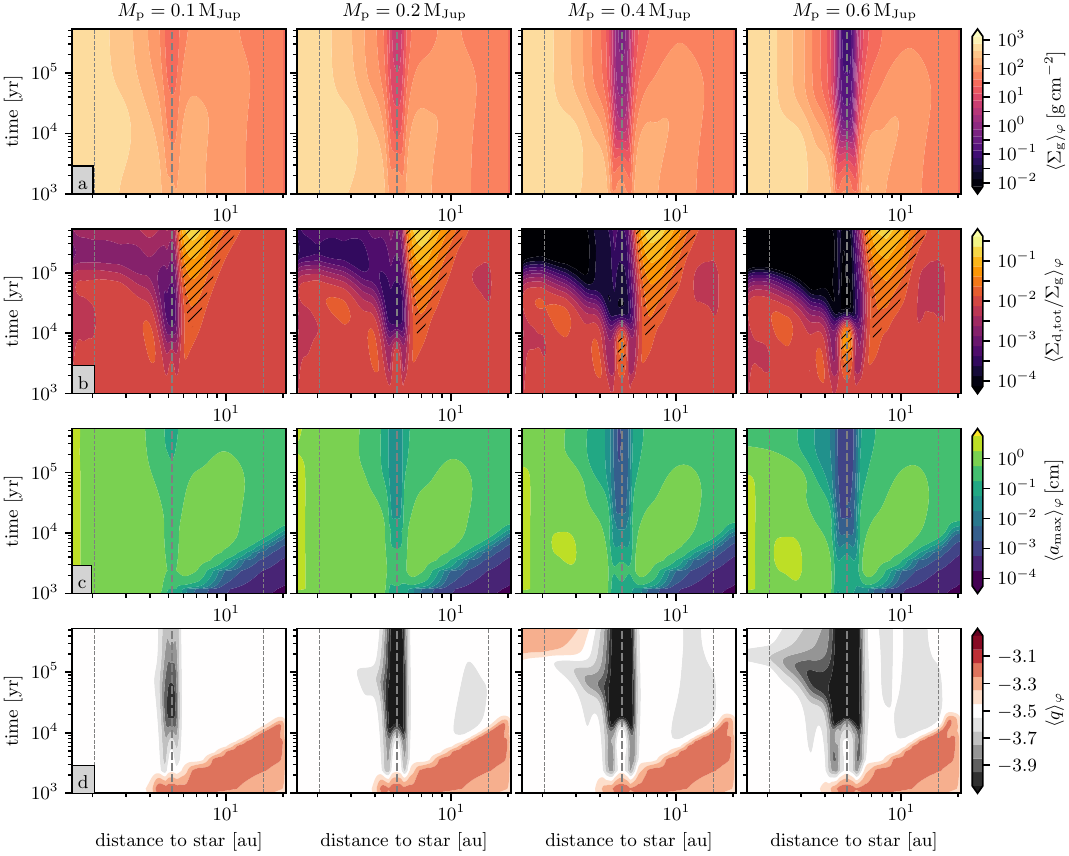}
    \caption{Time evolution of the azimuthally averaged disk structure of our simulations for different planetary masses (each column corresponds to a particular simulation; see titles). Row (a) shows the evolution of the gas density. Row (b) depicts the evolution of the dust-to-gas ratio. Hatched areas mark regions in which the \cite{Lim2024} criterion for strong clumping via the Streaming Instability is fulfilled.  Maximum particle sizes and size distribution power-law exponents are shown in rows (c) and (d) respectively. The thick dashed line marks the location of the planet. Thin dashed lines show the boundaries of the wave damping zones.}
    \label{fig:FullEvolMp}
\end{figure*}
Larger planets induce stronger torques onto the protoplanetary disk and thus carve deeper gaps. 
We investigate planetary masses from \SI{0.1}{\Mjup} to \SI{0.6}{\Mjup}.
\autoref{fig:CompMp} depicts the azimuthally averaged disk structure at the end of our four simulations (after \SI{45000}{} planetary orbits).
The \SI{0.1}{\Mjup} planet carves a shallow gap in the gas (panel a) with a density depletion factor of $\sim 2$, while the most massive planet reduces the density in the co-orbital region by a factor of $\sim 1000$.
Since the gap becomes wider for larger planetary masses, we also find that dust trapping occurs further away from the planet for larger masses, as can be seen in panel (b) of \autoref{fig:CompMp}. A wider gap results in a longer distance for the dust to diffuse through before it can reach the inner disk. The dust densities in the inner disk are thus severely depleted in the simulations with higher planetary masses.
The maximum particle sizes in and around the gap also change drastically with planetary mass (panel c). In the \SI{0.1}{\Mjup} case, \si{\milli\meter}-sized grains can still be found within the gap. At larger planetary mass, smaller particles are effectively removed from the gap. 
For the most massive planet, only \si{\micron}-sized grains can still pass through the gap.
Particle sizes in the inner parts of the disk are also influenced by gap filtering. In the cases in which gap filtering is inefficient and the dust densities remain high in the inner disk, we find that particle sizes closely follow the analytically calculated fragmentation limit \citep[thin dashed lines in panel c, see][for definition of $a_\mathrm{frag}$]{Birnstiel2012}. For larger planetary masses, where the inner disk is severely dust depleted, growth timescales become so long that the dust particles coming from the outer disk cannot grow to the fragmentation limit within the runtime of our simulations. Here, larger, initially present particles drift out of the domain and are replaced by smaller dust at much lower density.
The power-law exponents of the size distributions in and inside the gap also vary with planetary mass, which is shown in panel (d).
The efficient removal of large particles from the planetary gap flattens the size distribution ($q\approx -4$). 

\rev{
For the highest planetary mass ($M_\mathrm{p}=\SI{0.6}{\Mjup}$), where the densities are severely depleted in the inner disk, we find the growth time scale to be too long to reestablish an MRN distribution in the simulation's runtime. Instead, the inwards diffusing particles inherit the size distribution from within the gap, where fewer large grains are present ($q<-3.5$). The exponents are higher for the intermediate case of $M_\mathrm{p}=\SI{0.4}{\Mjup}$, where the particles are \revII{starting} to approach the fragmentation line but have not yet reached it in the simulation's runtime. The size distributions in this particular case are thus still in the sweep-up dominated regime, where large particles absorb the smaller grains without yet fragmenting in mutual collisions. This leads to a top heavy distribution, i.e., $q>-3.5$.
}
\rem{The inwards diffusion of small dust with a flat size distribution can be seen to also change the power-law exponent in the inner disk for the high-mass cases. This effect only occurs for the massive planets because of the severely reduced dust densities which result in much longer dust growth time scales.}

\autoref{fig:FullEvolMp} provides a detailed look at the temporal evolution of the disk structure in our simulations.
Rows (a) and (b) show the effect of the planet on the gas and dust densities. 
Gas densities far inside the planet's orbit are only slightly changed due to the planet-disk interactions. Dust densities however become strongly diluted due to the effect of the pressure trap at the gap's outer edge. 
In every case, dust-to-gas ratios in the pressure traps reach values suitable for strong clumping via the Streaming Instability \citep{Lim2024}, as shown by the hatched areas in row (b). This means planetesimal formation could commence at the outer gap edge. Note, however, that we calculate the criterion based on the total dust density and the maximum Stokes number of the distribution, which is likely not appropriate for a polydisperse dust size distribution. We do not explicitly model planetesimal formation in this study.
Row (c) shows the evolution of the maximum particle size throughout our simulation. 
The fragmentation limit is reached within \SI{e4}{\years} at the outer edge of our simulation domain in every simulation.
The effect of the planetary potential on the maximum particle size can be seen to set in as soon as the particles grow larger than a micrometer. 
Larger grains are aerodynamically removed from the planets co-orbital region. This effect is much stronger for larger planets. As soon as the gas gap reaches its final depth, also the maximum particle size reaches its new equilibrium value in the gap, based in the fluxes in and out of the gap. Coagulation is not of importance within the gaps carved by the larger planets' due to the reduced dust densities.
This also affects the maximum grain size interior to the gap. Particle sizes adjacent to the inner gap edge decrease once the drift of large particles from the outer disk is blocked by the pressure maximum. Only \si{\micron}-sized grains can pass through, which then re-coagulate in the inner disk.
This also affects the evolution of the size distribution power-law exponent, which is shown in row (d). During the initial growth phase, size distributions tend to be top heavy, as fragmentation is not yet occurring and large grains sweep-up smaller particles. This is visible as the red, triangular region in in row (d) at times $<\SI{e4}{\years}$. Once fragmentation and coagulation have reached an equilibrium state, power laws relax towards an MRN-like distribution with $q=-3.5$.
As soon as the gap starts to form, large grains are efficiently removed from the co-orbital region via aerodynamic drag. The power law becomes flattened. 
Particles following this flat power-law distribution diffuse into the inner disk, where they re-coagulate, thus gradually restoring an MRN distribution. This process can be significantly slowed down compared to the initial growth phase, as can be seen in the \SI{0.4}{\Mjup} and \SI{0.6}{\Mjup} simulations. Because the dust densities are strongly reduced in the inner disk in these cases, we find that the growth time scale also becomes long, which is why an MRN distribution is not restored in these simulations in the inner disk. 


\subsection{Fragmentation Velocity}\label{sec:vfrag}
\begin{figure}[ht]
    \centering
    \includegraphics[width=\linewidth]{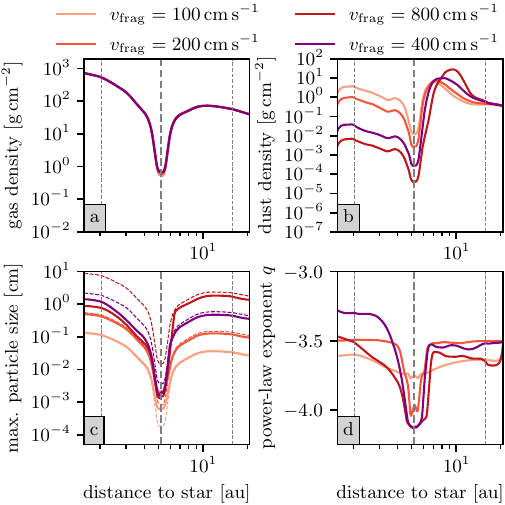}
    \caption{Azimuthally averaged disk profiles for the four simulations with different fragmentation velocities after \SI{45000}{} orbits ($\sim\SI{533602}{\years}$) of evolution. Panel (a) shows the radial gas density profiles, panel (b) depicts the dust density profiles. Maximum particle sizes of the power-law size distributions are shown in panel (c). Thin dashed lines mark the analytically estimated fragmentation limit. The local power-law exponents of the dust size distributions are shown in panel (d).}
    \label{fig:CompVfrag}
\end{figure}

The fragmentation velocity is the only parameter we vary that almost exclusively influences the dust component (except for minor changes in the gas structure due to backreaction, see panel (a) in \autoref{fig:CompVfrag}).
Panel (c) of \autoref{fig:CompVfrag} shows the maximum particle sizes in these simulations. For the smallest value of the fragmentation velocity that we investigate ($v_\mathrm{frag}=\SI{100}{\centi\meter\per\second}$), particles can grow to $\sim \SI{100}{\micron}$-sizes at the outer gap edge, whereas the largest value of  ($v_\mathrm{frag}=\SI{800}{\centi\meter\per\second}$) allows particles to grow $\sim$50 times as large. Particle sizes are significantly reduced in the planet's co-orbital region, where large grains are removed by aerodynamic drag. 
Dust density gaps are significantly deeper in simulations with larger fragmentation velocities in which dust accumulates more in the pressure trap.
Dust densities in the inner disk are thus strongly depleted for larger fragmentation velocities. A consequence of the low densities in the inner disk is that the particles' growth timescale becomes long. This means that the dust that reaches the inner disk does not reach the fragmentation limit in the simulations runtime for the fragmentation velocities \SI{400}{\centi\meter\per\second} and \SI{800}{\centi\meter\per\second} (see dashed lines in panel c, which show the analytically estimated fragmentation limit).

The maximum particle sizes also influence the power-law exponents that are reached in coagulation/fragmentation equilibrium in the inner disk, as can be seen in panel (d). The simulation with the largest particles has slightly shallower dust size distributions in the outer disk, since grain-grain collisions are more strongly influenced by relative drift \citep{Birnstiel2011}. Similarly, collisions in simulations with very small particles can be driven by the small eddy regime of turbulence, which also results in slightly shallower size distributions.

\rev{Similar to cases of higher planetary mass, dust collision rates in the inner disk are severely reduced in the inner disk for higher fragmentation velocities due to the reduced dust density. The trends in the size distribution exponents in the inner regions are thus comparable to the trend with planetary mass (see \autoref{sec:Mplanet}).

A detailed view of the time evolution of the simulations with different fragmentation velocities is provided in \autoref{sec:App1A}.}

\rem{The overall evolution of the disks with different fragmentation velocities is shown in \autoref{fig:FullEvolVfrag}. Row (a) depicts the azimuthally averaged gas densities, which evolve virtually identical in all simulations. The gaps reach their final depth after $\sim$\SI{20000}{\years}.
The dust density evolution, shown in row (b), however, varies drastically for the different fragmentation velocities. The relatively small particles in the simulation with $v_\mathrm{frag}=\SI{100}{\centi\meter\per\second}$ can cross the gap easily and the dust-to-gas ratio in the inner disk stays almost constant throughout the simulation. The pressure maximum nonetheless traps dust and larger particles ($\gtrsim\SI{10}{\micron}$) are removed from the co-orbital region of the planet (row c). 
Significant amounts of small grains can still traverse the gap for $v_\mathrm{frag}=\SI{200}{\centi\meter\per\second}$. The dust-to-gas ratio in the inner disk is, however, starting to deplete after \SI{e5}{\years} but still remains higher than $10^{-3}$ even at the end of our simulation.
Trapping starts to significantly influence the inner disk's mass budget for the fragmentation velocities larger than \SI{200}{\centi\meter\per\second}. Particles reach \si{\milli\meter} to \si{\centi\meter} sizes in this simulation, meaning that only $\sim\SI{5}{\percent}$ of the dust mass trapped outside the planetary gap is smaller than $\sim \SI{30}{\micron}$.
Dust densities drop sharply in the inner disk once the planetary gap is fully formed. 
Densities in the inner disk deplete faster for larger fragmentation velocities in general because particles of larger Stokes numbers drift faster. In the case of $v_\mathrm{frag}=\SI{800}{\centi\meter\per\second}$, the inner disk is mostly depleted of material after $\sim \SI{50000}{\years}$ whereas this process takes approximately \SI{e5}{\years} for $v_\mathrm{frag}=\SI{400}{\centi\meter\per\second}$. 

For the largest fragmentation velocity, conditions even become suitable for strong clumping via the Streaming Instability in the inner disk before the dust has drifted out of the domain (hatched areas in row b).

The slopes of the dust size distributions in the inner disk is also mostly influenced by the gap once trapping becomes efficient. Particles less than \SI{10}{\micro \meter} which follow a flat size distribution diffuse into the inner disk, where they start to slowly re-coagulate. This process takes longer in the simulations with larger fragmentation velocities because of the severly depleted dust densities.}

\subsection[Alpha Parameter]{$\alpha$ Parameter}\label{sec:alpha}
\begin{figure}[ht]
    \centering
    \includegraphics[width=\linewidth]{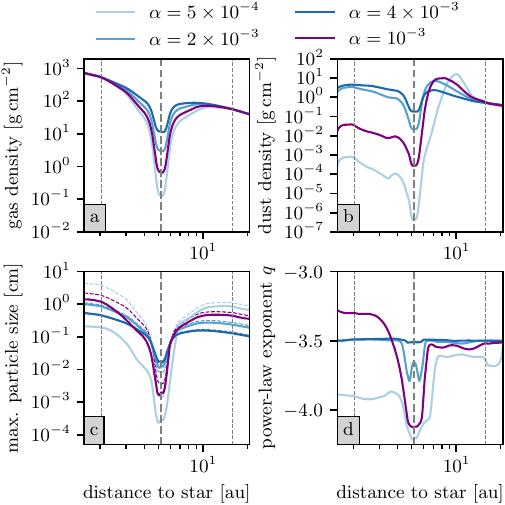}
    \caption{Azimuthally averaged disk profiles for the four simulations with different $\alpha$ parameters after \SI{45000}{} orbits ($\sim\SI{533602}{\years}$) of evolution. Panel (a) shows the radial gas density profiles, panel (b) depicts the dust density profiles. Maximum particle sizes of the power-law size distributions are shown in panel (c). Thin dashed lines mark the analytically estimated fragmentation limit. The local power-law exponents of the dust size distributions are shown in panel (d).}
    \label{fig:CompAlpha}
\end{figure}

The $\alpha$ parameter has by far the strongest influence on the simulation results in our study because we use the same $\alpha$ to prescribe the viscosity, turbulent particle collision velocities, and dust diffusivities.
Lower gas viscosities correspond to deeper gaps, as can be seen in panel (a) of \autoref{fig:CompAlpha}. The lowest value we investigate ($\alpha=5\times 10^{-4}$) results in a gap in which the gas densities are depleted by a factor of $10^3$.
Lower viscosities also result in larger maximum particle sizes \rev{because of the reduced turbulent collision speed. These larger particles} undergo faster radial drift and can be trapped more easily in pressure maxima. Maximum particle sizes scale approximately as $\propto \alpha^{-1}$, as can be seen in panel (c) of \autoref{fig:CompAlpha}.
Lower diffusivities furthermore mean that small grains need more time to diffuse through a gap.
In summary, small $\alpha$ result in larger particles, deeper gaps, and longer diffusion timescales, resulting in a severely depleted gap and inner disk in our simulation for $\alpha=5\times 10^{-4}$ (see panel b).
Lower $\alpha$ values also affect the size distribution exponent since relative drift velocities become more relevant in particle collisions. The simulations with low $\alpha$ therefore show size distributions which are generally shallower than in the high $\alpha$ cases (panel d). 

\rev{As for the case of different planetary masses and fragmentation velocities, size distributions exponents in the inner regions of the disk depend on the available dust mass budget, which determines the grain collision time scale (see \autoref{sec:Mplanet}). 

We also provide a detailed overlook of the time evolution of the simulations with different turbulence parameters in \autoref{sec:App1B}.}

\rem{\autoref{fig:FullEvolAlpha} gives an overview of the global disk evolution for the four simulations with different $\alpha$ parameters. The gas gaps are fully formed after $\sim\SIrange{2e4}{4e4}{\years}$ in all cases, as can be seen in row (a). In the $\alpha=5\times 10^{-4}$ case, we find that the inner disk becomes almost completely dust depleted after $\sim \SI{8e4}{\years}$ (row b). This process takes longer for $\alpha=10^{-3}$.
Larger $\alpha$ values result in shallower gaps, more diffusion, and smaller particles and the dust densities remain high for $\alpha=2\times 10^{-3}$ and $\alpha=4\times 10^{-3}$ in the inner disk.
In the $\alpha=5\times 10^{-4}$ case, particles crossing the gap cannot grow to the fragmentation limit in the inner disk within the simulation's runtime due to the highly depleted dust densities (see row c).
The maximum particle sizes in the inner disk remain slightly below the fragmentation limit for $\alpha=10^{-3}$ and the distributions are in coagulation-fragmentation equilibrium in the simulations with $\alpha>10^{-3}$. This can also be seen in the size distribution power-law exponents in row (d). Dust particles in the severely depleted inner disk for $\alpha=5\times 10^{-4}$ retain the shallow size distribution they inherit from the gap due to the very low dust-dust collision rates. For $\alpha=10^{-3}$ we can see that the size distribution has reached a state in which larger grains have started to sweep-up the smaller particles in the inner disk but have not yet reached the fragmentation limit. Therefore, the size distribution is slightly steeper than in the fragmentation limit.
At higher dust densities \st{($\alpha>10^{-3}$)}, particles traversing the gap quickly evolve towards fragmentation-coagulation equilibrium.}

\begin{figure*}[ht]
    \centering
    \includegraphics[width=\linewidth]{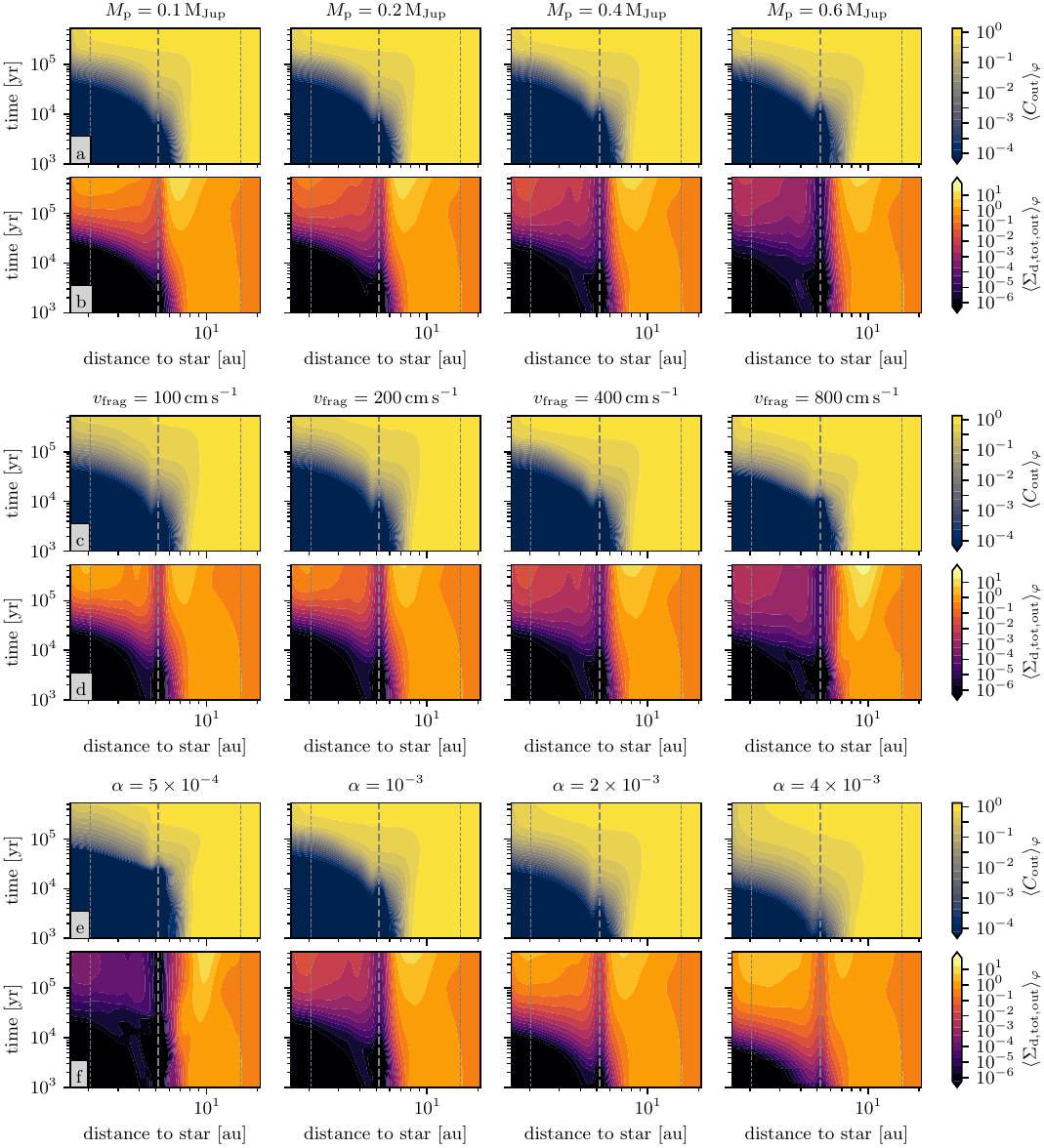}
    \caption{\rev{Concentrations (rows a, c, e) and column densities (row b, d, f) of dust originating from the outer disk in our simulations with different planetary masses (rows a and b), fragmentation velocities (rows c and d) and turbulence parameters (rows e and f). The thick dashed line marks the location of the planet. Thin dashed lines show the boundaries of the wave damping zones.}}
    \label{fig:ContamAll}
\end{figure*}

\subsection{Compositional Evolution}\label{sec:CompositionalEvolution}
\begin{figure}[ht]
    \centering
    \includegraphics[width=\linewidth]{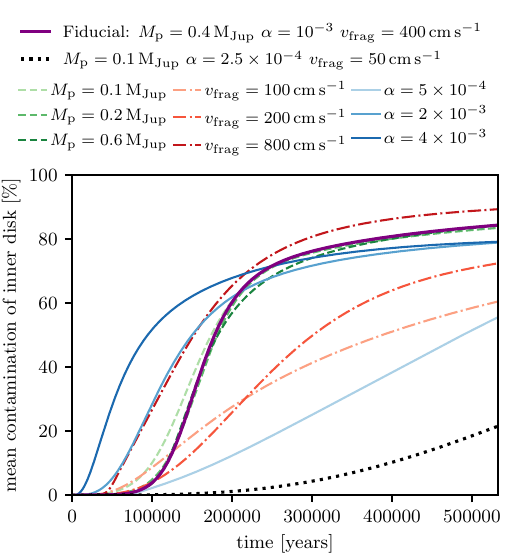}
    \caption{Average concentration of dust in the inner disk that originates from the outer disk as a function of time for the different simulations presented in this study.}
    \label{fig:Cont}
\end{figure}

\begin{figure}[ht]
    \centering
    \includegraphics[width=\linewidth]{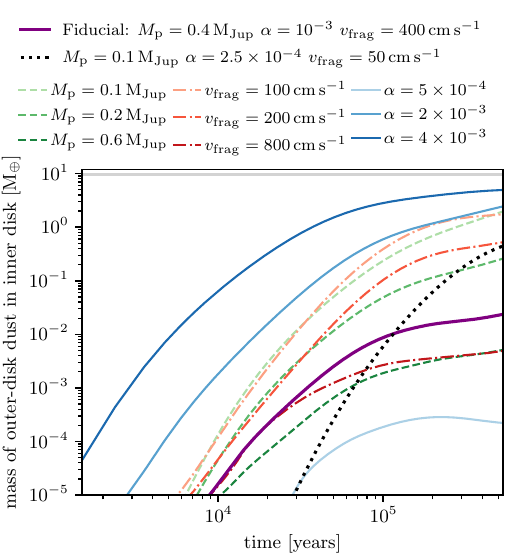}
    \caption{Total mass of dust within the planet's orbital radius coming from the outer disk ($R>\SI{7.8}{\AU}$) as a function of time for the different simulation runs. The horizontal gray line shows the total dust mass in the inner disk at the beginning of the simulation ($\sim\SI{9}{\Mearth}$).}
    \label{fig:Mint}
\end{figure}

Previous works \citep{Weber2018, Stammler2023} have tracked the amount of dust that reaches in inner disk by setting the dust densities in the inner disk to zero, thus allowing them to exactly measure the mass that reaches the inner disk.
Since we want to model the entire disk and also keep track of the composition of the material, we are instead following the evolution of two passive scalars, representing the local concentration of outer-disk material (see \autoref{sec:PassiveScalars}).
\autoref{fig:ContamAll} shows the time evolution of the azimuthally averaged concentration of outer-disk material and the respective dust surface densities 
\begin{align}
    \langle C_\mathrm{out}\rangle_\varphi &= \frac{1}{2\pi} \int_0^{2\pi}\frac{ C_0\rhodsm + C_1\rhodlr}{\rhodsm + \rhodlr} \,\mathrm{d}\varphi \\
    \langle \Sigma_\mathrm{d,tot,out}\rangle_\varphi &= \frac{1}{2\pi} \int_0^{2\pi} C_0\rhodsm + C_1\rhodlr \, \mathrm{d}\varphi \, ,
\end{align}
for the simulations with four different planetary masses, fragmentation velocities, and $\alpha$ parameters respectively.
Concentrations start to tend towards outer-disk composition after $\sim\SI{e5}{\years}$ in all \rev{cases}. 
At this time, most of the originally present dust in the inner disk has drifted out of the simulation domain and been replaced by inwards diffusing/drifting grains from the outer disk. Averaged over the inner disk, we find that \SI{80}{\percent} of the remaining dust has outer-disk composition for all planetary masses (see green dashed lines in \autoref{fig:Cont}).
The actual density of outer-disk material varies strongly across the simulations. 
We measure the mass of outer-disk dust that reaches the inner disk as a function of time
\begin{equation}
    M_\mathrm{d,out\rightarrow in} = \int_{R_\mathrm{in}}^{R_\mathrm{p}} \int_0^{2\pi} R\, \Sigma_\mathrm{d,tot,out} \, \mathrm{d}\varphi\,  \mathrm{d}R\, ,
\end{equation}
shown in \autoref{fig:Mint}.
Most mass reaches the inner disk in the simulation with the shallowest gap.
None of the presented simulations can maintain a complete separation between the inner and outer disk.
For the \SI{0.6}{\Mjup} planet, only \SI{5e-3}{\Mearth} reach the inner disk. In the case of the \SI{0.1}{\Mjup}, however, which corresponds to roughly the mass of Jupiter's core \citep{Wahl2017}, \SI{1.96}{\Mearth} of outer-disk material reach the inner disk (see green dashed lines in \autoref{fig:Mint}). 
This mass corresponds to $\sim\SI{20}{\percent}$ of the initially present dust mass in the inner disk. 

Similar trends in mass contamination can be seen in the simulations for different fragmentation velocities \rev{(\autoref{fig:ContamAll}, rows c and d)}. 
Even for a planetary mass of $\SI{0.4}{\Mjup}\approx \SI{6.4}{\Mth}$, we find that \SI{1.73}{\Mearth} of outer-disk material can reach the inner disk within \SI{5e5}{\years} if the fragmentation velocity is \SI{100}{\centi\meter\per\second}.
Although the mass of material reaching the inner disk is similar to the case of the \SI{0.1}{\Mjup} planet at \SI{400}{\centi\meter\per\second}, we find that the concentration of outer-disk material is quite different in both runs.
While the inner disk is composed to \SI{82.5}{\percent} of outer disk material in the case of the \SI{0.1}{\Mjup} planet with a dust fragmentation velocity of \SI{400}{\centi\meter\per\second}, we find that the inner disk is composed to only \SI{62.1}{\percent} of outer-disk material for the \SI{0.4}{\Mjup} planet with \SI{100}{\centi\meter\per\second} fragmentation velocity.
The reason for this is the higher drift speed of the dust in the inner disk for the larger fragmentation velocity case. Another reason \rev{is} the higher dust density in the pressure trap for larger fragmentation velocities due to the increased drift speed and trapping efficiency. Density gradients are thus \rev{steeper} and the resulting mass fluxes into the gap larger for higher fragmentation velocities.
Since more of the original material is lost due to drift in the case of $v_\mathrm{frag}=\SI{400}{\centi\meter\per\second},\, M_\mathrm{p}=\SI{0.1}{\Mjup}$, we find higher concentrations of outer-disk dust here. 
The composition of planetesimals formed in the inner disk, therefore not only depends on the mass that can traverse the gap, but also in the drift speed of the dust in the inner disk, which determines the amount of material with inner-disk composition, which is still present when planetesimal formation can commence.

Compositional changes of the inner disk do not vary much with planetary mass because the ratio of outer-disk material to inner disk material is mostly determines by the depletion timescale of the inner disk due to drift. Diffusion timescales of outer-disk dust into the inner disk are of similar length and almost independent of planetary mass.
Compositional changes already set in before the bulk of the mass has reached the inner disk, as dust is drifting away faster than it is replenished via diffusion/drift from the outer disk.

As discussed before, $\alpha$ is by far the most influential parameter in our simulations because gap depth, particle sizes, and diffusion speed scale with it.
Compositional changes thus vary strongly with $\alpha$. At $\alpha>10^{-3}$, we can see concentrations shifting towards \SI{70}{\percent} outer-disk composition within just \SI{3e5}{\years}. An average composition of \SI{50}{\percent} outer-disk material is reached within $\sim \SI{e5}{\years}$ for $\alpha=4\times 10^{-3}$ and within $\sim\SI{1.5e5}{\years}$ for $\alpha=2\times 10^{-3}$.
Similarly, significant amounts of mass reach the inner disk at large $\alpha$ values. We find \SI{4.95}{\Mearth} masses of outer-disk material within the inner disk at the end of the simulation for the case of $\alpha=4\times 10^{-3}$ and \SI{2.44}{\Mearth} in the simulation with
$\alpha=2\times 10^{-3}$. 
The situation changes drastically for our simulation with $\alpha=5\times 10^{-4}$. Here, the planet acts as an effective barrier against the inwards drifting particles.
While the composition changes towards \SI{50}{\percent} outer-disk material after \SI{5e5}{\years}, only \SI{2e-4}{\Mearth} of dust make it into the inner disk during the runtime of our simulation.
Dust densities are thus severely depleted in the inner disk for this setup.
Any planetesimals formed within the planetary radius, would thus have almost pristine inner-disk composition.


\subsection{Low Viscosity, Low Fragmentation Velocity}\label{sec:FringeCase}
\begin{figure}[ht]
    \centering
    \includegraphics[width=\linewidth]{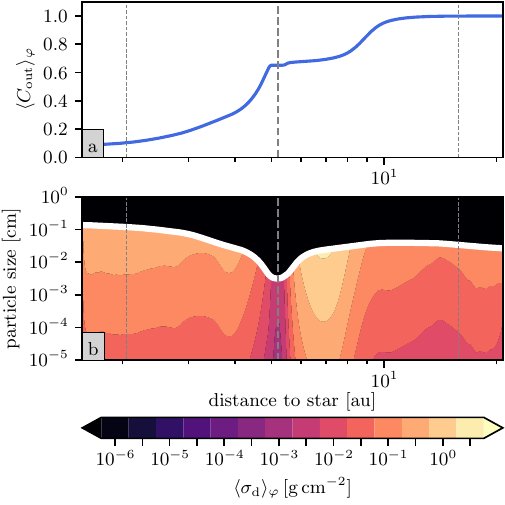}
    \caption{Azimuthally averaged concentration of outer-disk dust (a) and dust size distribution (b) after \SI{45000}{} planetary orbits for the low viscosity ($\alpha=2.5\times 10^{-4}$) and low fragmentation velocity ($v_\mathrm{frag}=\SI{50}{\centi\meter\per\second}$) setup with a planetary mass roughly corresponding to Jupiter's core ($M_\mathrm{p}=\SI{0.1}{\Mjup} =\SI{31.8}{\Mearth}$).}
    \label{fig:FullDistr_Fringe}
\end{figure}

Here, we investigate a more extreme parameter combination that has not been studied in the previous sections: a low viscosity ($\alpha=2.5\times 10^{-4}$) and low fragmentation velocity ($v_\mathrm{frag}=\SI{50}{\centi\meter\per\second}$) setup with a planetary mass roughly corresponding to the upper limit for the mass of Jupiter's core \citep[$M_\mathrm{p}=\SI{0.1}{\Mjup} =\SI{31.8}{\Mearth}$, see][]{Wahl2017}.
Albeit their small fragmentation velocities, particles can still grow to \si{\milli\meter} sizes because of the low turbulent gas velocities. The size distributions are mostly drift-fragmentation-limited in this scenario, leading to a distribution exponent of $\sim-3.75$. Using \autoref{eq:estimate} with $\amax=\SI{1}{\milli\meter}$, we can estimate that $\sim\SI{24}{\percent}$ of the dust mass in the pressure trap is composed of particles smaller than \SI{10}{\micron}---a significantly higher percentage than in the fiducial simulation. 
The comparably low Stokes numbers in this scenario result in a much longer drift timescale of the particles, which means the inner disk becomes only slowly depleted. Dust from the outer disk indeed has enough time to reach the inner disk before it gets depleted by radial dust drift. The lasting presence of the original grain population in the inner disk means that the average concentration of outer-disk material remains relatively low during the runtime of our simulation. After \SI{45000}{} orbits, only \SI{20}{\percent} of the dust in the inner disk originates from the outer region (\autoref{fig:Cont}). 
The local concentration of outer-disk dust is also shown in panel (a) of \autoref{fig:FullDistr_Fringe}. 
The total mass of outer-disk dust in the inner disk, however, is large (see \autoref{fig:Mint}). Approximately \SI{0.44}{\Mearth} of dust with outer-disk composition can be found in the inner disk. This amount could in fact significantly influence the composition of planetesimals formed after $\sim\SI{e5}{\years}$, depending on how close to the planetary gap they form.

\section{Discussion} \label{sec:discussion}
Our simulations show that the evolution of local dust size distributions due to coagulation and fragmentation cannot be ignored in studies of gap filtration and dust composition in protoplanetary disks (see also \autoref{sec:AppB}).
We have probed four different fragmentation velocities---all within the currently discussed ranges derived from laboratory experiments and simulations \citep{Gundlach2015}---in our parameter study. The outcomes are vastly different in terms of dust distribution and composition in the inner disk. While \SI{1.73}{\Mearth} of dust from the outer disk are present in the inner disk for $v_\mathrm{frag}=\SI{100}{\centi\meter\per\second}$ at the end of the simulation, we find that gap filtration becomes orders of magnitude more effective for $v_\mathrm{frag}=\SI{800}{\centi\meter\per\second}$, where only \SI{5e-3}{\Mearth} can be found in the inner disk.

Within the relevant mass range for giant planet cores and forming gas giants, investigated here, filtration can only be highly efficient if diffusivities and viscosities are very low or if fragmentation velocities are high.
Large fragmentation velocities of cold water ice (as present at the disk regions investigates here) have been ruled out in recent laboratory studies, which found enhanced stickiness only for temperatures close to the evaporation temperature \citep{Musiolik2019}. Values in the range of \SI{100}{\centi\meter\per\second} are generally to be expected. However, large uncertainties exist and the exact values of the fragmentation velocity remains uncertain.
Viscosities might have been low at the site of Jupiter's formation, due to its location within the dead zone \citep[e.g.,][]{Okuzumi2012}. We have furthermore not investigated the effects of magnetized disk winds \citep[e.g.,][]{Blandford1982} or photoevaporation \citep[e.g.,][]{Clarke2001} on the dust.

Some amount of contamination of the inner disk by outer-disk material is always to be expected if coagulation/fragmentation are considered. 
\rev{Given these findings, Jupiter could have only maintained an isotopic dichotomy in the early Solar system if very specific conditions were met. To maintain a separation, Jupiter must have grown very fast, since an object with a mass of its core could not have stopped the dust from reaching the inner disk.
Alternative explanations for the isolated formation of the CC and NC chondrite reservoirs might thus be necessary. 
One possibility, suggested by \cite{Brasser2020} could be the early partitioning of the disk into rings before giant planet formation. Substructures such as rings and vortices could have been formed by late-stage infall \citep{Kuznetsova2022}, and might thus also be linked to the actual origins of the nucleosynthetic dichotomy itself \citep{Nanne2019}.
Planetesimal formation at the iceline \citep{Drazkowska2013} could have resulted in the emergence of two spatially and temporally separated populations of meteorite parent bodies in the early Solar system without Jupiter's presence \citep{Lichtenberg2021}.}

Compared to previous two-dimensional studies \citep{Zhu2012, Weber2018, Drazkowska2019}, we have investigated disks at lower $\alpha$ values and found that significant leakage is still possible in these cases if the fragmentation velocity is low.
Simulations at low viscosities and/or high planetary masses remain challenging even in two-dimensional setups because of the extensive number of planetary orbits that have to be simulated in order to reach an equilibrium of coagulation, fragmentation, drift and diffusion. 

In accordance with previous works, we showed that filtration is more efficient for larger particles. In contrast to past studies, however, we included the effect of dust coagulation and fragmentation. The presence of large particles in our simulations is thus always leading to the presence of small particles through fragmentation. This supply of small grains at the gap edge thus increases the leakage and is continuously replenished by the inwards drifting large grains that collect in the pressure trap and fragment there. These findings are in agreement with one-dimensional studies, using full coagulation prescriptions \citep[e.g.,][]{Stammler2023, Homma2024}, which generally find some degree of leakage.
With respect to the isotopic dichotomy in the Solar System, our studies strongly suggest that alternative explanations are likely necessary to explain the spatial and temporal separation of the NC and CC chondrite-forming mass reservoirs. Jupiter's core could have only maintained this separation if very specific conditions were met at the time of chondrite formation.

The highly variable filtration efficiency demonstrated in our study has implications for the composition of the inner disk, the central star, and planetesimal formation in the inner regions in general.

The concept of the pebble isolation mass \citep{Morbidelli2012, Lambrechts2014,Bitsch2018,Ataiee2018} assumes that the accretion of solids onto a forming planet is stopped once its mass becomes large enough to open a gap and stop the pebble flux.
In light of previous studies with dust coagulation \citep[e.g.,][]{Stammler2023, Drazkowska2019,Homma2024}, our results confirm that this is likely oversimplified if coagulation and fragmentation occur. 
Smaller grains are able to traverse the gap and might still reach the planet, thus contributing to the envelope's opacity and dust mass.
Although the total mass gain through this process might be limited (depending on the diffusivities), cooling times could be changed, thus impacting the accretion of gas.
We do not model this process in our simulations since we cannot resolve the circumplanetary disk. \rem{Furthermore, three-dimensional flows likely alter the pebble mass that can be accreted by a planet compared to two-dimensional setups.}
\rev{
Models of pebble accretion \citep[e.g.,][]{Ormel2010, Lambrechts2012, Lambrechts2014a, DePaula2019} and runaway gas accretion in two dimensional simulations \citep[e.g.,][]{Kley1999, DAngelo2008, DePaula2019a, Rometsch2024} rely on assumptions on the three-dimensional gas and dust structure and the accretion rate's dependence on the distance to the planet. Three-dimensional simulations have shown that the accretion flows are inherently three-dimensional as well, also following meridional trajectories \citep{Bi2021, Bi2023, VanClepper2025, Huang2025}. Two-dimensional models of pebble accretion onto planets might nonetheless be able to approximate the results of three-dimensional simulations, as shown by \cite{Picogna2018}. We have included a single example of a simulation with dust accretion in \autoref{sec:AppE}.
Generally speaking, dust and gas accretion will alter the results of our parameter study to some degree. At the same time, they will also alter the investigated parameters themselves (i.e., the planetary mass). All of our planet's could potentially accrete gas in a runaway process and therefore could not be treated as constant mass objects in a realistic simulation. We leave the task of incorporating these effects for future investigations.
}

It has been theorized by \cite{Booth2020} that the depletion of metals in Solar twins could be a sign for giant planet formation where the resulting gap prevents dust from accreting onto the central star, thus, reducing its metallicity. Our results show that this might only be possible for specific parameter combinations that allow the gap to filter dust efficiently. 

The almost complete depletion of dust in the inner disk in some of our simulations furthermore might be in tension with continuum observations of younger protoplanetary disks. While transition disk are characterized by inner cavities \citep[see, e.g.][]{Francis2020}, most younger systems still exhibit continuum flux from the inner regions \citep[e.g.,][]{Andrews2018}.
This can be interpreted in two ways: Either a number of dust traps exist in the inner disk that conserve the primordial dust population in place, or particles from the outer disk are able to traverse the depleted gaps as seen in many of our simulations.
Recently, \cite{Sierra2025} observed that leaky dust traps also exist in PDS-70.

Ultimately, three-dimensional simulations \citep[as in][]{Huang2025} must be coupled to dust coagulation models to gain a more realistic, and quantitative evaluation of the gap filtration efficiency. 

The dust mass present in the inner disk regions of our simulations is not equal to the mass that will be converted into planetesimals, as dust is constantly drifting and diffusing in through the gap and out of the inner regions through the inner domain boundary. If planetesimals form in the inner disk, e.g., via the cold-finger effect and dust trapping at the water ice line \citep{Drazkowska2013}, masses and compositions will differ from the amount of dust present at a single simulation snapshot, as presented here.
Since dust in the inner disk drifts quickly for large fragmentation velocities, it is possible that a first generation of planetesimals can be formed at the water ice line \citep{Lichtenberg2021}, before dust from the outer disk had enough time to diffuse and drift there.
Future studies must include a planetesimal formation mechanism if conclusions on the actual composition and isotopic anomalies of chondrites are to be drawn.

Planetesimal formation can furthermore influence the dust mass budget in the pressure trap \citep{Stammler2019}. We have shown in \Cref{fig:FullEvolMp,fig:FullEvolVfrag,fig:FullEvolAlpha}, that the \cite{Lim2024} criterion for strong clumping via the Streaming Instability is met at the outer gap edge in most of our simulations. If dust is efficiently converted into larger bodies there, the respective depletion of dust in the trap could reduce the flux of small grains through the gap.
Whether the \cite{Lim2024} criterion also holds for a polydisperse size dispersion is unclear. Here, we have used the total dust density and the maximum particle size to compute the critical metallicity for strong clumping in a turbulent protoplanetary disk.
We will study the impact of gap filtration on planetesimal composition and formation, and vice versa in a future series of simulations.
\begin{figure*}[ht]
    \centering
    \includegraphics[width=\linewidth]{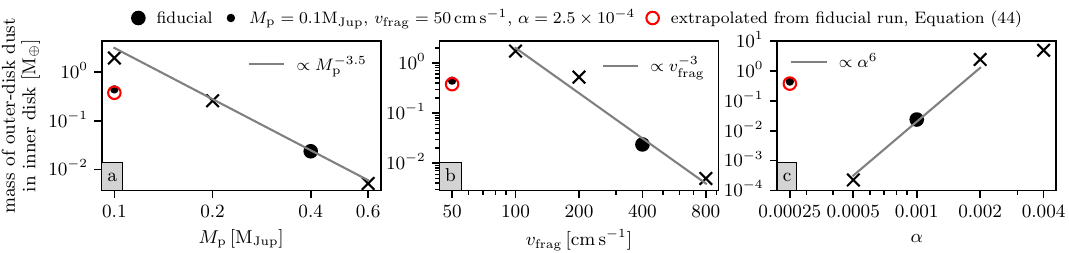}
    \caption{Mass of outer-disk dust at the end of the simulations (after \SI{45000}{} orbits) as a function of the varied parameters. Power-laws are plotted as gray lines to give a rough idea of the scaling and are not fitted to the data.}
    \label{fig:results}
\end{figure*}

\section{Summary}\label{sec:summary}
In this work, we have conducted a small parameter study for two-dimensional simulations of planet-disk systems, including dust coagulation \revII{and fragmentation} with the \tpop{} method. 
We were interested in the inner disk's mass budget and its composition in terms of the mass concentration of material originating from the outer disk. 
Our models include the effects of dust-gas backreaction, coagulation/fragmentation, and dust concentration diffusion.
All simulations resulted in some degree of\, ``contamination'' of the inner disk by outer-disk dust at varying degrees, depending on the simulation parameters. 
While the dust mass budget in the inner disk can be severely depleted for large planetary masses, fragmentation velocities and small $\alpha$ parameters, we find that whatever amount of material is still present after \SI{5e5}{\years}, it will be composed to at least \SI{50}{\percent} of outer-disk material for the given parameter combinations. The low-viscosity, low-fragmentation velocity case is the exception, since the original dust population is still largely present in the inner disk. Significant amounts of material still reach the inner disk in this scenario but are more strongly diluted in the existing dust reservoir. This resulted in a concentration of $\sim \SI{20}{\percent}$ of outer-disk dust. 

Despite these significant compositional changes throughout the inner disk, planetesimals might still be formed predominantly from material of inner-disk composition for certain systems. This will be true for low $\alpha$ cases, large fragmentation velocities and large planetary masses, where filtering is efficient and the inner disk becomes severely depleted once the dust has drifted away.  
Although we don't directly model planetesimal formation, we can give upper limits for the dust mass that is available at a given time. 
If $\alpha$ is large ($\gtrsim 2\times 10^{-3}$), or if the fragmentation velocity is low ($\lesssim\SI{400}{\meter\per\second}$), or if the planetary mass is low ($\lesssim\SI{0.2}{\Mjup}$), we find that $\sim\SIrange{0.1}{4}{\Mearth}$ of outer-disk material can be present in the inner disk after \SI{5e5}{\years} of evolution.

If, however, we deal with a low viscosity disk in which the dust has a high fragmentation velocity, dust filtration might be highly efficient if a planetary core can grow fast enough to carve a deep gap. 
The filtration efficiency scales strongly with all the varied parameters (see \autoref{fig:results}). While leakage is highly reduced for larger planetary masses ($M_\mathrm{d,out\rightarrow in}\propto M_\mathrm{p}^{-3.5}$, panel a) and high fragmentation velocities ($M_\mathrm{d,out\rightarrow in}\propto v_\mathrm{frag}^{-3}$, panel b), we find that the by far most influential parameter is the viscosity/diffusivity ($M_\mathrm{d,out\rightarrow in}\propto\alpha^6$ panel c). This strong scaling is the result of $\alpha$'s threefold effect in our simulations. It determines the gap's depth, the fragmentation-limited maximum particle size, and the diffusivity of the dust.
At very high viscosities, the scaling seems to flatten for the presented parameter combinations. Unfortunately, $\alpha$ is also one of the least constrained parameters in our simulations and might vary by orders of magnitude depending on the physical conditions within a protoplanetary disk.

While varying individual parameters can give us an idea of the general scaling laws of the problem, certain parameter combinations can still lead to significant leakage, even if one of the parameters lies in the unfavorable regime for leakage.
To investigate the outcome of a low-viscosity, low-fragmentation-velocity disk with Jupiter's core at \SI{5.2}{\AU}, we ran a model with $\alpha=2.5\times 10^{-4}$, $v_\mathrm{frag}=\SI{50}{\centi\meter\per\second}$ and $M_\mathrm{p}=\SI{0.1}{\Mjup}$, roughly corresponding to the most massive estimate for the core mass. We found that, despite the very low diffusivity, significant amounts of dust can still make it into the inner disk in this scenario due to the low fragmentation velocity and shallow gap. In this case, we measure \SI{0.444}{\Mearth} of outer-disk dust in the region interior to the planet at the end of the simulation.

We test the combined scalings from \autoref{fig:results} on this result. Given the mass of outer-disk dust in the inner disk from the fiducial simulation, we use
\begin{align} \label{eq:scaling}
    M_\mathrm{d,out\rightarrow in}^{\SI{45000}{orbits}} \approx \SI{0.0236}{\Mearth}&\times\left(\frac{M_\mathrm{p}}{\SI{0.4}{\Mjup}}\right)^{-3.5}\\ \nonumber
    &\times \left(\frac{v_\mathrm{frag}}{\SI{400}{\centi\meter\per\second}}\right)^{-3} \\ \nonumber
    &\times\left(\frac{\alpha}{10^{-3}}\right)^{6}\,
\end{align}
to estimate the final mass of outer-disk dust in the inner disk for $\alpha=2.5\times 10^{-4}$, $v_\mathrm{frag}=\SI{50}{\centi\meter\per\second}$ and $M_\mathrm{p}=\SI{0.1}{\Mjup}$.
This expression gives \SI{0.378}{\Mearth} compared to the actual value of \SI{0.444}{\Mearth} (i.e., an \SI{18}{\percent} deviation), shown as red circles in \autoref{fig:results}. The effects of the individual parameters thus appear to be separable. Thoroughly testing this relation will require more simulations and we leave this task for future studies. Since the simulation with $\alpha=4\times 10^{-3}$ deviates from the general trend in $\alpha$ (panel c in \autoref{fig:results}), we expect \autoref{eq:scaling} to not hold for high viscosities, where the gap is very shallow and the inflow of material through the gap approaches the accretion rate in a smooth disk.

Given these strong scaling relations and the resulting uncertainties, we conclude that a clear compositional dichotomy in the Solar System could have only been preserved by an early-formed, fast growing Jupiter in a disk with very low turbulence, in which the dust particles don't fragment at collision velocities of up to several \SI{100}{\centi \meter \per\second}. Our results suggest that alternative explanations for the preservation of the isotopic dichotomy and other filtering related processes should be considered.
A mass reservoir of purely inner-disk composition might be formed in any case if chondrites and planetesimals form via some mechanism within a few \SI{e4}{\years} in the inner disk, before outer-disk material has diffused through the gap. Whether this is possible remains unclear and will be the subject of future studies.

\section*{Acknowledgments}
The Flatiron Institute is supported by the Simons Foundation. 
T.P. would like to thank Til Birnstiel, Aida Behmard, Joanna Dr{\k{a}}{\.z}kowska, Linn Eriksson, and Alexandros Ziampras for helpful discussions on the topic.

\bibliographystyle{aasjournalv7}
\bibliography{Literature.bib,PhilLiterature.bib}

\appendix

\section[Global disk evolution for different alpha and vfrag]{Global disk evolution for different turbulence parameters and fragmentation velocities}
\rev{
\subsection{Fragmentation velocities}
\begin{figure*}[b]
    \centering
    \includegraphics[width=\linewidth]{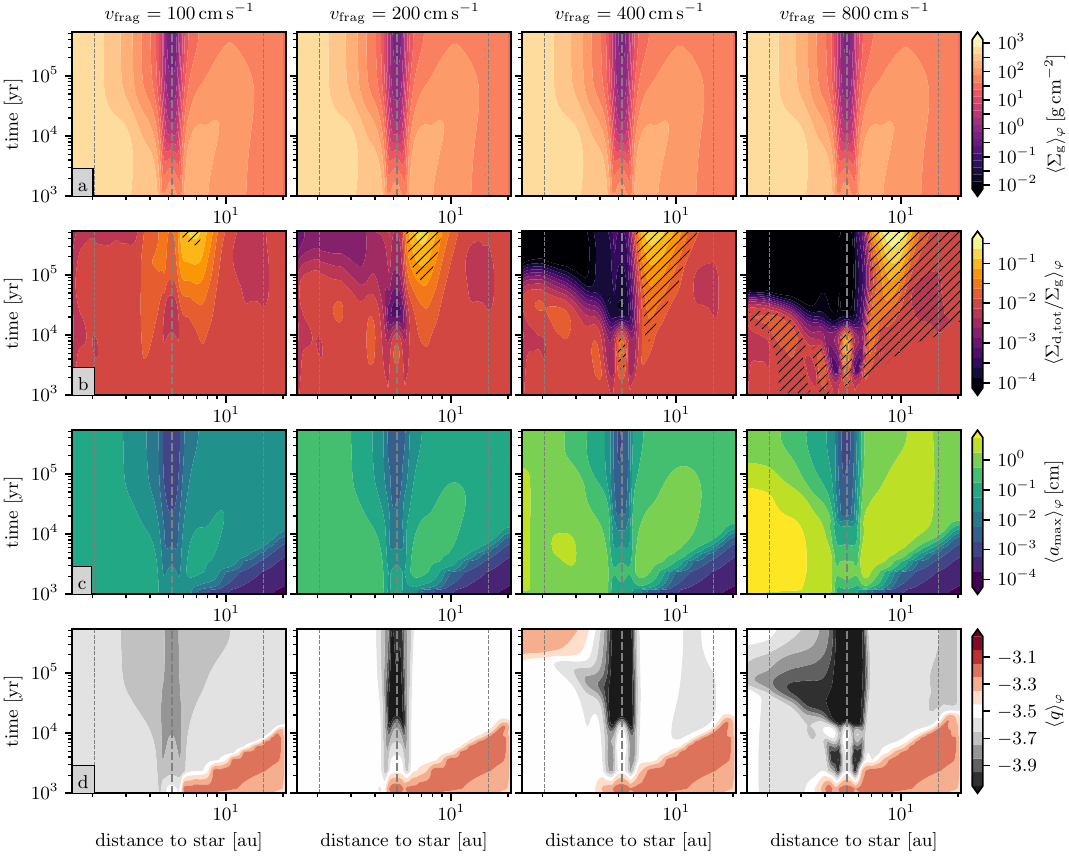}
    \caption{Time evolution of the azimuthally averaged disk structure of our simulations for different fragmentation velocities (each column corresponds to a particular simulation; see titles). Row (a) shows the evolution of the gas density. Row (b) depicts the evolution of the dust-to-gas ratio. Hatched areas mark regions in which the \cite{Lim2024} criterion for strong clumping via the Streaming Instability is fulfilled. Maximum particle sizes and size distribution power-law exponents are shown in rows (c) and (d) respectively. The thick dashed line marks the location of the planet. Thin dashed lines show the boundaries of the wave damping zones.}
    \label{fig:FullEvolVfrag}
\end{figure*}
\label{sec:App1A}
The overall evolution of the disks with different fragmentation velocities is shown in \autoref{fig:FullEvolVfrag}. Row (a) depicts the azimuthally averaged gas densities, which evolve virtually identical in all simulations. The gaps reach their final depth after $\sim$\SI{20000}{\years}.
The dust density evolution, shown in row (b), however, varies drastically for the different fragmentation velocities. The relatively small particles in the simulation with $v_\mathrm{frag}=\SI{100}{\centi\meter\per\second}$ can cross the gap easily and the dust-to-gas ratio in the inner disk stays almost constant throughout the simulation. The pressure maximum nonetheless traps dust and larger particles ($\gtrsim\SI{10}{\micron}$) are removed from the co-orbital region of the planet (row c). 
Significant amounts of small grains can still traverse the gap for $v_\mathrm{frag}=\SI{200}{\centi\meter\per\second}$. The dust-to-gas ratio in the inner disk is, however, starting to deplete after \SI{e5}{\years} but still remains higher than $10^{-3}$ even at the end of our simulation.
Trapping starts to significantly influence the inner disk's mass budget for the fragmentation velocities larger than \SI{200}{\centi\meter\per\second}. Particles reach \si{\milli\meter} to \si{\centi\meter} sizes in this simulation, meaning that only $\sim\SI{5}{\percent}$ of the dust mass trapped outside the planetary gap is smaller than $\sim \SI{30}{\micron}$.
Dust densities drop sharply in the inner disk once the planetary gap is fully formed. 
Densities in the inner disk deplete faster for larger fragmentation velocities in general because particles of larger Stokes numbers drift faster. In the case of $v_\mathrm{frag}=\SI{800}{\centi\meter\per\second}$, the inner disk is mostly depleted of material after $\sim \SI{50000}{\years}$ whereas this process takes approximately \SI{e5}{\years} for $v_\mathrm{frag}=\SI{400}{\centi\meter\per\second}$. 

For the largest fragmentation velocity, conditions even become suitable for strong clumping via the Streaming Instability in the inner disk before the dust has drifted out of the domain (hatched areas in row b).

The slopes of the dust size distributions in the inner disk is also mostly influenced by the gap once trapping becomes efficient. Particles less than \SI{10}{\micro \meter} which follow a flat size distribution diffuse into the inner disk, where they start to slowly re-coagulate. This process takes longer in the simulations with larger fragmentation velocities because of the severely depleted dust densities.
}

\rev{
\subsection{Turbulence parameters}
\label{sec:App1B}
\begin{figure*}[ht]
    \centering
    \includegraphics[width=\linewidth]{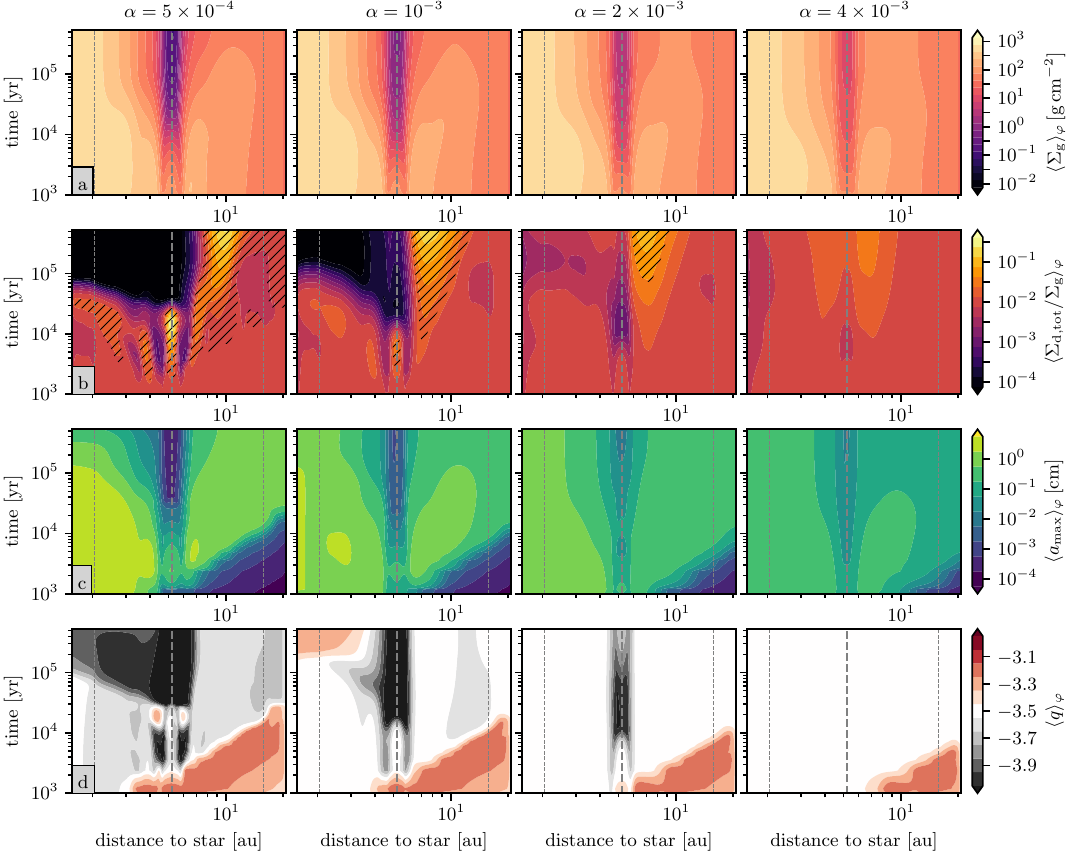}
    \caption{Time evolution of the azimuthally averaged disk structure of our simulations for different $\alpha$ parameters (each column corresponds to a particular simulation; see titles). Row (a) shows the evolution of the gas density. Row (b) depicts the evolution of the dust-to-gas ratio. Hatched areas mark regions in which the \cite{Lim2024} criterion for strong clumping via the Streaming Instability is fulfilled. Maximum particle sizes and size distribution power-law exponents are shown in rows (c) and (d) respectively. The thick dashed line marks the location of the planet. Thin dashed lines show the boundaries of the wave damping zones.}
    \label{fig:FullEvolAlpha}
\end{figure*}
\autoref{fig:FullEvolAlpha} gives an overview of the global disk evolution for the four simulations with different $\alpha$ parameters. The gas gaps are fully formed after $\sim\SIrange{2e4}{4e4}{\years}$ in all cases, as can be seen in row (a). In the $\alpha=5\times 10^{-4}$ case, we find that the inner disk becomes almost completely dust depleted after $\sim \SI{8e4}{\years}$ (row b). This process takes longer for $\alpha=10^{-3}$.
Larger $\alpha$ values result in shallower gaps, more diffusion, and smaller particles and the dust densities remain high for $\alpha=2\times 10^{-3}$ and $\alpha=4\times 10^{-3}$ in the inner disk.
In the $\alpha=5\times 10^{-4}$ case, particles crossing the gap cannot grow to the fragmentation limit in the inner disk within the simulation's runtime due to the highly depleted dust densities (see row c).
The maximum particle sizes in the inner disk remain slightly below the fragmentation limit for $\alpha=10^{-3}$ and the distributions are in coagulation-fragmentation equilibrium in the simulations with $\alpha>10^{-3}$. This can also be seen in the size distribution power-law exponents in row (d). Dust particles in the severely depleted inner disk for $\alpha=5\times 10^{-4}$ retain the shallow size distribution they inherit from the gap due to the very low dust-dust collision rates. For $\alpha=10^{-3}$ we can see that the size distribution has reached a state in which larger grains have started to sweep-up the smaller particles in the inner disk but have not yet reached the fragmentation limit. Therefore, the size distribution is slightly steeper than in the fragmentation limit.
At higher dust densities, \rev{i.e., in the high viscosity cases where larger amounts of dust diffuse into the inner disk} \st{($\alpha>10^{-3}$)}, particles traversing the gap quickly evolve towards coagulation\rev{-fragmentation} equilibrium.
}
\clearpage
\section{Global dust size distributions}
\begin{figure*}[ht]
    \centering
    \includegraphics[width=\linewidth]{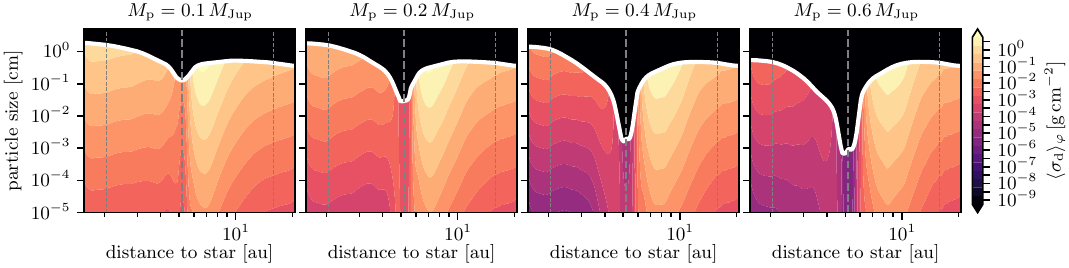}
    \caption{Global evolution of the grain size distributions (azimuthally averaged) for our simulations with four different planetary masses. Thin dashed lines indicate the boundaries of the radial damping zones. The thick dashed line indicates the location of the planet.}
    \label{fig:SizeDistrMplanet}
\end{figure*}
\begin{figure*}[ht]
    \centering
    \includegraphics[width=\linewidth]{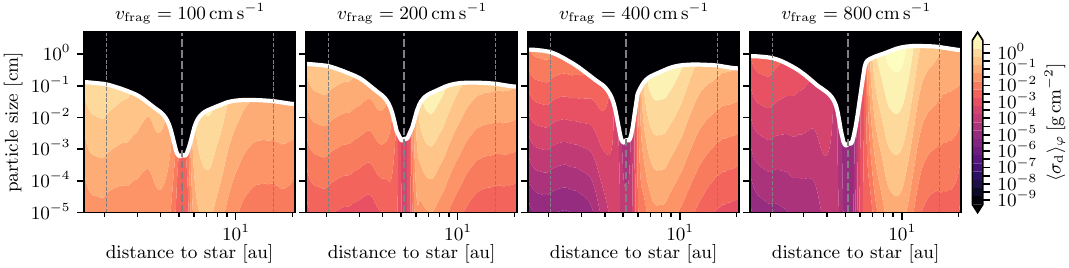}
    \caption{Global evolution of the grain size distributions (azimuthally averaged) for our simulations with four different fragmentation. Thin dashed lines indicate the boundaries of the radial damping zones. The thick dashed line indicates the location of the planet.}
    \label{fig:SizeDistrvfrag}
\end{figure*}
\begin{figure*}[ht]
    \centering
    \includegraphics[width=\linewidth]{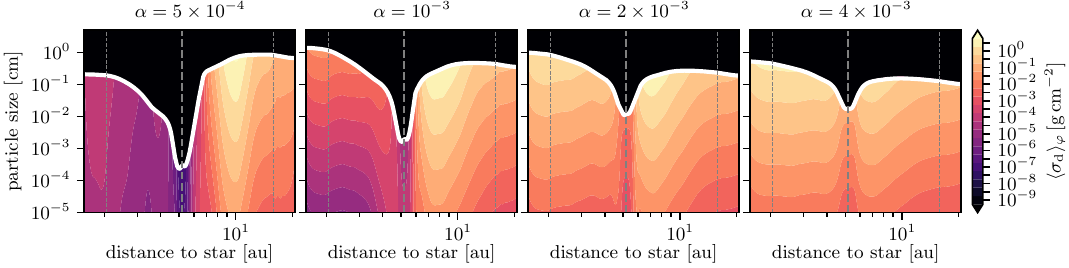}
    \caption{Global evolution of the grain size distributions (azimuthally averaged) for our simulations with four different $\alpha$ parameters. Thin dashed lines indicate the boundaries of the radial damping zones. The thick dashed line indicates the location of the planet.}
    \label{fig:SizeDistrAlpha}
\end{figure*}
Here we provide the global dust size distribution plots for all simulations performed in this study. \autoref{fig:SizeDistrMplanet} shows the azimuthally averaged dust size distributions for the four different planetary masses, each with $\alpha=10^{-3}$ and $v_\mathrm{frag}=\SI{400}{\centi\meter\per\second}$. The runs for different $v_\mathrm{frag}$ are presented in \autoref{fig:SizeDistrvfrag}. The planetary mass is kept at \SI{0.4}{\Mjup}, and $\alpha=10^{-3}$.
The simulations with different $\alpha$ viscosities are shown in \autoref{fig:SizeDistrAlpha}, where we kept $v_\mathrm{frag}=\SI{400}{\centi\meter\per\second}$ and $M_\mathrm{p}=\SI{0.4}{\Mjup}$.

\section{Comparison to a fixed-particle-size simulation}
\label{sec:AppB}
\begin{figure*}[ht]
    \centering
    \includegraphics[width=\linewidth]{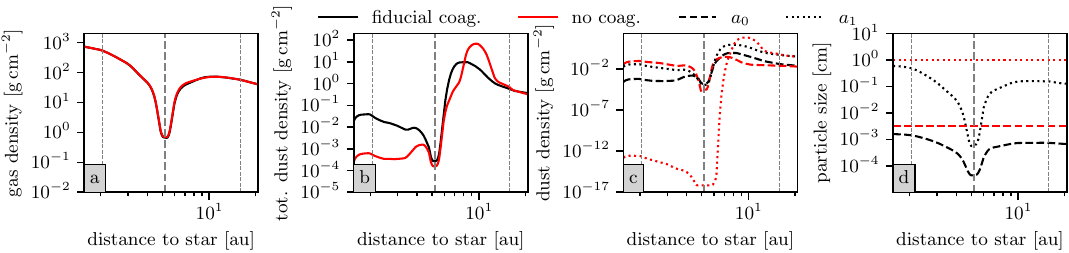}
    \caption{Comparison between the fiducial simulation ($v_\mathrm{frag}=\SI{400}{\cm\per\second}$, $\alpha=10^{-3}$, $M_\mathrm{p}=\SI{0.4}{\Mjup} \approx \SI{6.4}{\Mth}$) with dust \revII{coagulation and fragmentation} and a simulation with fixed particle sizes and no mass exchange between the two dust fluids after \SI{45000}{} planetary orbits. Panel (a) shows the gas densities, panel (b) and (c) depict the total dust densities and the two components respectively. We show the mass-averaged particle sizes for the coagulation simulation (black) and the constant particle sizes from the comparison run in panel d.}
    \label{fig:Fixed}
\end{figure*}
Here, we provide a single comparison between our fiducial simulation ($v_\mathrm{frag}=\SI{400}{\cm\per\second}$, $\alpha=10^{-3}$, $M_\mathrm{p}=\SI{0.4}{\Mjup} \approx \SI{6.4}{\Mth}$) and a fixed-particle-size simulation. We initialize the simulation without coagulation with an MRN distribution made of 2 dust fluids and a maximum particle size of $a_1 = \SI{1}{\centi\meter}$. The particle Stokes numbers for the two dust fluids are then constantly given by $a_1$ and $\sqrt{a_1\amin}$ for the entire simulation, where $\amin=\SI{0.1}{\micron}$.
The resulting disk structure after \SI{45000}{} planetary orbits is shown in \autoref{fig:Fixed}. Panel (a) depicts the gas densities which are basically identical in both simulations. The total dust density (panel b) and the two individual densities (panel c), however differ drastically in the simulation without coagulation\revII{ and fragmentation}. The larger dust fluid in the simulation without \revII{collisional dust evolution} ($a_1 = \SI{1}{\centi\meter}$) is completely blocked at the outer gap edge. Within the gap, dust densities generally fall to the floor value and in the inner disk remain at negligible values. The smaller grains ($a_0=\sqrt{a_1\amin}$), however, can diffuse through the gap and densities are thus still higher in the inner disk. 
Mass exchange between the two population in the simulation with coagulation\revII{ and fragmentation} constantly replenished the amount of large grains in the inner disk and small grains in the pressure trap, leading to a completely different density distribution. In fact, the density of large grains is 10 orders of magnitude higher in the inner disk if \revII{fragmentation} is considered.
The total dust densities profile differs drastically from the \tpop{} result. The main density maximum, mostly composed of large grains, is at \SI{10}{\AU}, whereas the fiducial run has a density maximum at $\sim \SI{8}{\AU}$.
Similar to the comparisons shown in \cite{Drazkowska2019}, this highlights the importance of considering dust size distributions and the coagulation and fragmentation process in multifluid hydrodynamic simulations of protoplanetary disks. 

\section{Effects of spiral density waves on the dust size distribution}
\begin{figure*}[ht]
    \centering
    \includegraphics[width=\linewidth]{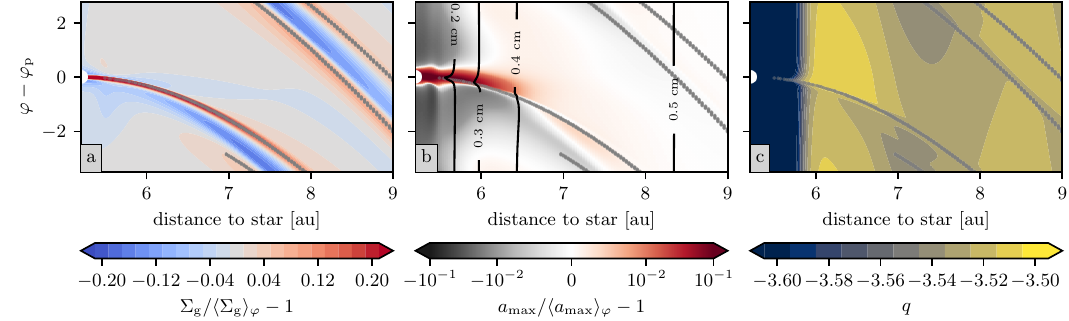}
    \caption{Zoomed-in snapshot of our simulation with $M_\mathrm{p}=\SI{0.6}{\Mjup}$, $\alpha=10^{-3}$, and $v_\mathrm{frag}=\SI{400}{\centi\meter\per\second}$.} 
    \label{fig:spiral_waves}
\end{figure*}

Recently, \cite{Eriksson2025} investigated the possible effects of spiral density waves on dust fragmentation in local planet-disk simulations via post-processed particle trajectories. They found that particle collision velocities can be significantly enhanced in the vicinity of spiral density waves, above the point of grain fragmentation.
We are only having a look at our simulation with the most massive planet with \SI{0.6}{\Mjup} here to investigate if we can observe a similar effect.
\autoref{fig:spiral_waves} depicts a zoomed-in snapshot of this simulation after \SI{30000}{} planetary orbits of evolution. Panel (a) shows the local gas density perturbations with respect to the azimuthally averaged gas density, which we used as a tracer for the location of the spiral density waves. By applying a peak finding algorithm to every azimuthal slice of the domain, we obtained the gray contours shown in all panels of \autoref{fig:spiral_waves}.
We find a \SI{20}{\percent} density enhancement at the spiral wave in the vicinity of the planet, which is located at the white marker in all panels.
The effect on the maximum particle size is shown in panel (b), which depicts relative deviation of the maximum particle size with respect to the azimuthal average.
It can be seen that the particles trailing the spiral density waves are in fact a few \si{\percent} smaller than the particles leading the waves. Furthermore, large dust seems to follow the flow towards the planet.
Drift-dominated collisions become more prevalent within the wave, as predicted by \cite{Eriksson2025}. The effect is, however, very weak, probably because the particles only spend a small fraction of an orbit in the region of steep radial pressure gradients which are caused by the wave.
Effects on the size distribution are visible in panel (c), which shows the size distribution exponents. The dust in the area trailing the spiral density wave has a slightly shallower size distribution, meaning that the amount of small grains is enhanced due to the stronger effect of collision caused by relative drift. 
These effects become significantly weaker with distance to the planet.

In accordance with \cite{Drazkowska2019}, we confirm that spiral density waves have a small effect on the dust size distributions close to a massive planet. The resulting changes are, however, likely of no major consequence for the filtering efficiency of the gap, or planetesimal formation in the pressure maximum for this specific parameter combination.

\rev{
\section{Importance of planetary accretion}\label{sec:AppE}
We have not included the effect of pebble accretion onto the planet in our simulations. Removing dust from the planet's pebble accretion radius could reduce the influx of dust into the inner disk and thus alter our results. 
We present here a first look on the possible magnitude of this effect.
To check whether mass accretion onto the planet significantly alters our findings we performed a single simulation (fiducial simulation parameters), in which we remove dust from the vicinity of the planet.
For this we reduce the dust density within the planet's Hill sphere on the free-fall timescale 
\begin{equation}
    \tau_\mathrm{ff}=\frac{\pi}{2}\sqrt{\frac{R_\mathrm{acc}^3}{2GM_\mathrm{p}}},
\end{equation}
where $R_\mathrm{acc}=a_\mathrm{H}$ in our simulation.
To avoid unsteady behavior at the Hill radius, we continuously ramp up the accretion rate, using a smooth transition function at $d<a_\mathrm{H}$, where $d$ is the distance to the planet
\begin{equation}
    f_\mathrm{acc} = \frac{1}{2}\cos{\left(\frac{d}{R_\mathrm{acc}}\right)} + \frac{1}{2}.
\end{equation}
The sink term is then defined as
\begin{equation}
    \frac{\mathrm{d}\Sigma_\mathrm{d,0/1}}{\mathrm{d}t} = -\frac{f_\mathrm{acc}\Sigma_\mathrm{d,0/1}}{\tau_\mathrm{ff}},
\end{equation}
which means we subtract the accreted density from the conserved variable as 
\begin{equation}
\Delta \Sigma_\mathrm{d,0/1} = \Sigma_\mathrm{d,0/1}\left[1 - \exp\left(-\frac{f_\mathrm{acc}\, \mathrm{d}t}{\tau_\mathrm{ff}}\right)\right],
\end{equation}
where $\Sigma_\mathrm{d,0/1}$ is the column density from the previous timestep. We employ the same prescription to the dust momenta.
This assumes particles are efficiently accreted once they enter the planet's Hill sphere. 

Whether pebbles are accreted or not depends on whether they have entered the accretion radius corresponding to their Stokes number. This might either be in the Bondi or Hill regime \citep{Lambrechts2014}.  We do not make this distinction here.

\begin{figure}
    \centering
    \includegraphics[width=0.5\linewidth]{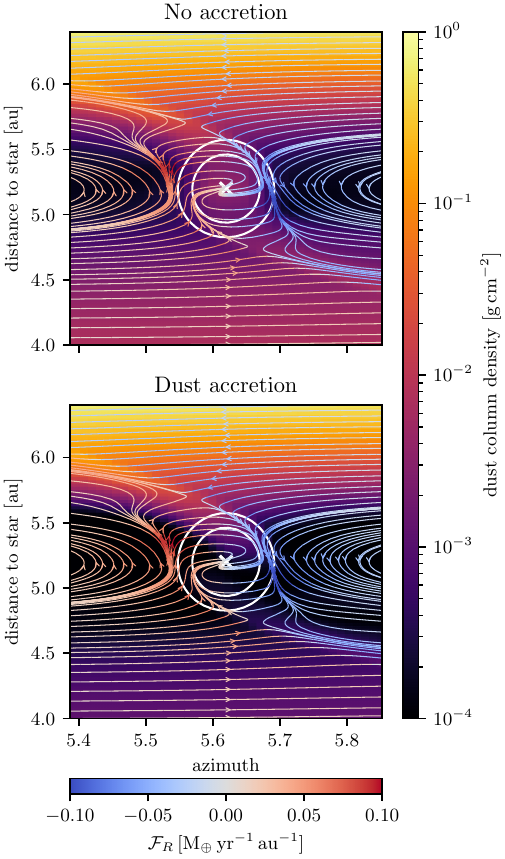}
    \caption{\rev{Comparison between the streamlines in the circumplanetary region of our fiducial simulation compared to the simulation with dust accretion.}}
    \label{fig:StreamlinesCompare}
\end{figure}
We show the resulting flow pattern, compared to the flow pattern in our fiducial simulation in \autoref{fig:StreamlinesCompare}. As can be seen, the streamlines are not significantly altered by the dust accretion prescription. Mass, however, is removed from the flow and the resulting contamination of the inner disk is thus reduced. 
\begin{figure}
    \centering
    \includegraphics[width=0.5\linewidth]{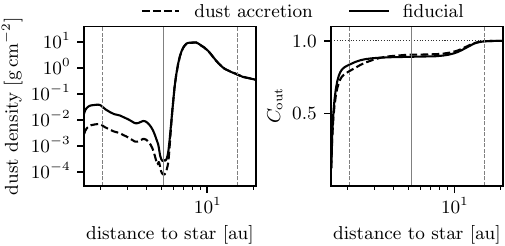}
    \caption{\rev{Comparison between the fiducial simulation (without dust accretion, solid lines) and a simulation with dust accretion (dashed lines).}}
    \label{fig:AccCompare}
\end{figure}
We compare the fiducial simulation to the simulation with dust accretion in \autoref{fig:AccCompare}. Dust densities in the inner disk are reduced by a factor of $\sim 5$ in the inner disk if our dust accretion prescription is employed. The concentration of outer disk material, however is not significantly changed because most of the primordial dust population has drifted out of the inner domain.
How this behavior changes with the investigated parameters should be further investigated in a future study. To ensure a self-consistent treatment of dust and gas accretion, one has to employ a three-dimensional simulation which also has to include gas accretion and a consistently growing planetary mass. 
This further highlights the importance of two-dimensional and three-dimensional hydrodynamic simulations compared to one-dimensional disk models.
}

\revII{
\section{Characteristic particle sizes throughout the disk}
Here, we present the characteristic particle sizes at the end of our fiducial simulation to demonstrate that only micron-sized particles can traverse the gap and reach the inner disk. \autoref{fig:OverviewSize} shows the three characteristic particle sizes of our size distributions, i.e., the maximum particle size at which the power-law distribution is truncated and the mass-averaged particle sizes of the two dust populations, which define their Stokes numbers and thus determine the dust dynamics. It can be seen that within the gap, mostly particles in the range of \SIrange{e-4}{e-3}{\centi\meter} (purple colors) can be found, while the flows around the planet also contain slightly larger particles (green colors). The shallow power-law exponent in the gap region, including the direct vicinity of the planet, means that almost half of the dust mass is contained in the small population in this case. 
The large particles that have reached the fragmentation limit (blue colors in the first panel) can not penetrate the gap. The entire flux into the inner region is thus dominated by the small particle size end of the size distribution, which is constantly replenished by fragmenting larger grains.
\begin{figure*}[ht]
    \centering
    \includegraphics[width=\linewidth]{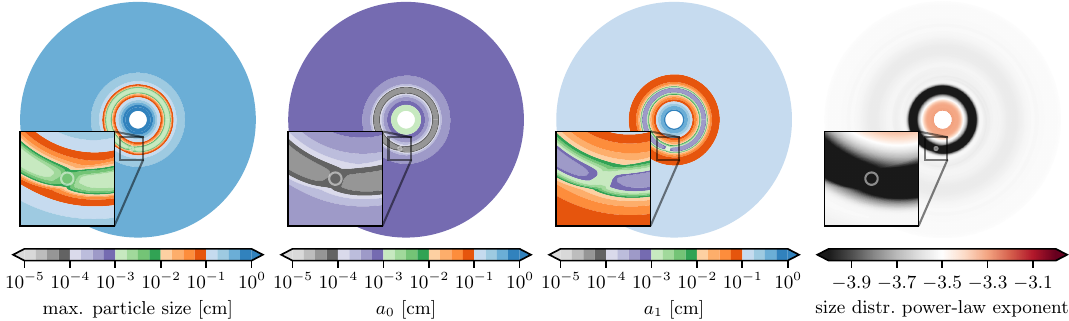}
    \caption{\revII{Characteristic particle sizes of the \tpop{} size distribution and power-law exponent of the distribution at the end of our fiducial simulation. From left to right: maximum particle size, mass-averaged particle size of the small population, mass-averaged particle size of the large population, power-law exponent.}}
    \label{fig:OverviewSize}
\end{figure*}

}

\end{CJK*}
\end{document}